\documentclass[a4paper]{raa}

\usepackage{graphicx,times}             
\usepackage{natbib}
\usepackage{amssymb,amsmath}
\usepackage{color}
\usepackage{ulem}
\usepackage{threeparttable}
\usepackage{adjustbox}
\usepackage{multirow}
\usepackage[figuresright]{rotating}
\bibpunct{(}{)}{;}{a}{}{,}
\begin{document}
   \title{LAMOST MRS-N Observations of the W80 Region
}

   \volnopage{Vol.0 (20xx) No.0, 000--000}      
   \setcounter{page}{1}          
\author{Yao Li
      \inst{1,2}
   \and Chao-Jian Wu
      \inst{1,3}
   \and Yong-Qiang Yao
     \inst{1}
   \and Wei Zhang
      \inst{1,3}
   \and Jia Yin
      \inst{1,2}
     \and Juan-Juan Ren
      \inst{1,4}
   \and Chih-Hao Hsia
      \inst{5}
   \and Rui Zhuang
      \inst{6}
   \and Jian-Jun Chen
      \inst{1,3}
   \and Yu-Zhong Wu
      \inst{1,3}
   \and Hui Zhu
      \inst{3}
   \and Bin Li
      \inst{7,8}
   \and Yong-Hui Hou
      \inst{2,9}
   \and Meng-Yuan Yao
	  \inst{10}
   \and Hong Wu
      \inst{1,3}
   }

\institute{{ National Astronomical Observatories, Chinese Academy of Sciences, Beijing 100101, China; \it chjwu@bao.ac.cn; \it yqyao@bao.ac.cn;}\\
        \and            
	  School of Astronomy and Space Science, University of Chinese Academy of Sciences, Beijing 100049, China; \\
        \and
	CAS Key Laboratory of Optical Astronomy, National Astronomical Observatories, Chinese Academy of Sciences, Beijing 100101, China;\\
	\and
	CAS Key Laboratory of Space Astronomy and Technology, National Astronomical Observatories, Chinese Academy of Sciences, Beijing 100101, China;\\
	\and
	State Key Laboratory of Lunar and Planetary Sciences, Macau University of Science and Technology, Taipa, Macau, China;\\
	\and
     University of Science and Technology Beijing, Beijing 100083, China;\\
	\and
	Purple Mountain Observatory, Chinese Academy of Sciences, Nanjing 210008, China;\\
	\and
	University of Science and Technology of China, Hefei 230026, China;\\
	\and
	Nanjing Institute of Astronomical Optics, Technology, National Astronomical Observatories, Chinese Academy of Sciences, Nanjing 210042, China;\\
	\and
	China Academy of Aerospace Electronics Technology, Beijing 100094, China;\\
   \vs\no
   {\small Received~~20xx month day; accepted~~20xx~~month day}}

\abstract{The spectral observations and analysis for the W80 Region are presented by using the data of Medium-Resolution Spectroscopic Survey of Nebulae (MRS-N) with the Large Sky Area Multi-Object Fiber Spectroscopy Telescope (LAMOST). 
A total of 2982 high-quality nebular spectra have been obtained in the 20 square degree field of view (FoV) which covers the W80 complex, and the largest sample of spectral data have been established for the first time. 
The relative intensities, radial velocities (RVs), and Full Widths at Half Maximum (FWHMs) are measured with the high spectral resolution of LAMOST MRS, for H$\alpha$  $\lambda$ 6563 \AA, [\ion{N}{ii}] $\lambda$$\lambda$ 6548 \AA, 6584 \AA \ , and [\ion{S}{ii}] $\lambda$$\lambda$ 6716 \AA, 6731 \AA \ emission lines. 
In the field of view of whole W80 Region, the strongest line emissions are found to be consistent with the bright nebulae, NGC 7000, IC 5070, and LBN 391, and weak line emissions also truly exist in the Middle Region, where no bright nebulae are detected by the wide-band optical observations. 
The large-scale spectral observations to the W80 Region reveal the systematic spatial variations of RVs and FWHMs, and several unique structural features. 
A `curved feature' to the east of the NGC 7000, and a `jet feature' to the west of the LBN 391 are detected to be showing with larger radial velocities. A `wider FWHM region' is identified in the eastern part of the NGC 7000. The variations of [\ion{S}{ii}] / H${\alpha}$ ratios display a gradient from southwest to northeast in the NGC 7000 region, and manifest a ring shape around the `W80 bubble' ionized by an O-type star in the L935. Further spectral and multi-band observations are guaranteed to investigate in detail the structural features.
}
\authorrunning{Y. Li,  $\&$ C. J. Wu }  
 \titlerunning{MRS-N Observations of the W80}  

\maketitle

%
%
\section{Introduction}           
\label{sect:intro}

Ionized hydrogen is generated when the ultraviolet light emitted by a massive star ionizes the surrounding gas. Previous work has shown that about 14\% to 22\% of the new generation of massive young stars are produced by the interaction of ionized hydrogen regions with surrounding molecular clouds \citep{2002AJ....124.3336H}. The study of ionized hydrogen is of great significance to our understanding of the interactions between massive stars and the surrounding molecular clouds.

The W80 complex is a classical \ion{H}{ii} region located in the Cygnus region, the nebular complex contains the North America Nebula (NGC 7000) and the Pelican Nebula (IC 5070)  (hereafter NAP nebulae), and a large dust cloud L935 \citep{1958BAN....14..215W, 2007ApJ...656..248C}. There is an  O-type star, 2MASS J20555125+4352246, embedded in the region as the ionizing source \citep{2005a&a...430..541c}. 
The central ionizing star is surrounded by a bubble with a radius of 20 pc, and the distance of the NAP region is about 800 pc \citep{2020a&a...633a..51z}.  The mass of the NAP nebulae is measured to be from  3$\times 10^4 M_{\odot}$ to 5$\times 10^4 M_{\odot}$ \citep{1993a&as..100..287f, 1980ApJ...239..121B}. T-Tauri star associations \citep{1958apj...128..259h} and a large number of outflow activities have been observed in the NAP Nebulae, such as the Herbig-Haro objects(HH objects), the molecular hydrogen emission-line objects (MHOs), and the H$_{2}$O masers \citep{2003AJ....126..893B,2011A&A...528A.125A}.
The molecular gas spectra, like NH$_{3}$, $^{12}$CO, $^{13}$CO, C$^{18}$O, and HCO$^{+}$, have been observed towards the L935 dust cloud \citep{2014AJ....147...46Z}. Using the Spitzer MIPS data, \cite{2009apj...697..787g} have identified more than 2,000 candidates of young stellar objects (YSOs) inside the NAP nebulae. Thus the W80 complex is one of the ideal targets to study the interactions of ionized hydrogen regions with molecular clouds in the process of star formation and early evolution.

\begin{figure}[!htp]
  \centering
  \includegraphics[width=0.7\textwidth,angle=0]{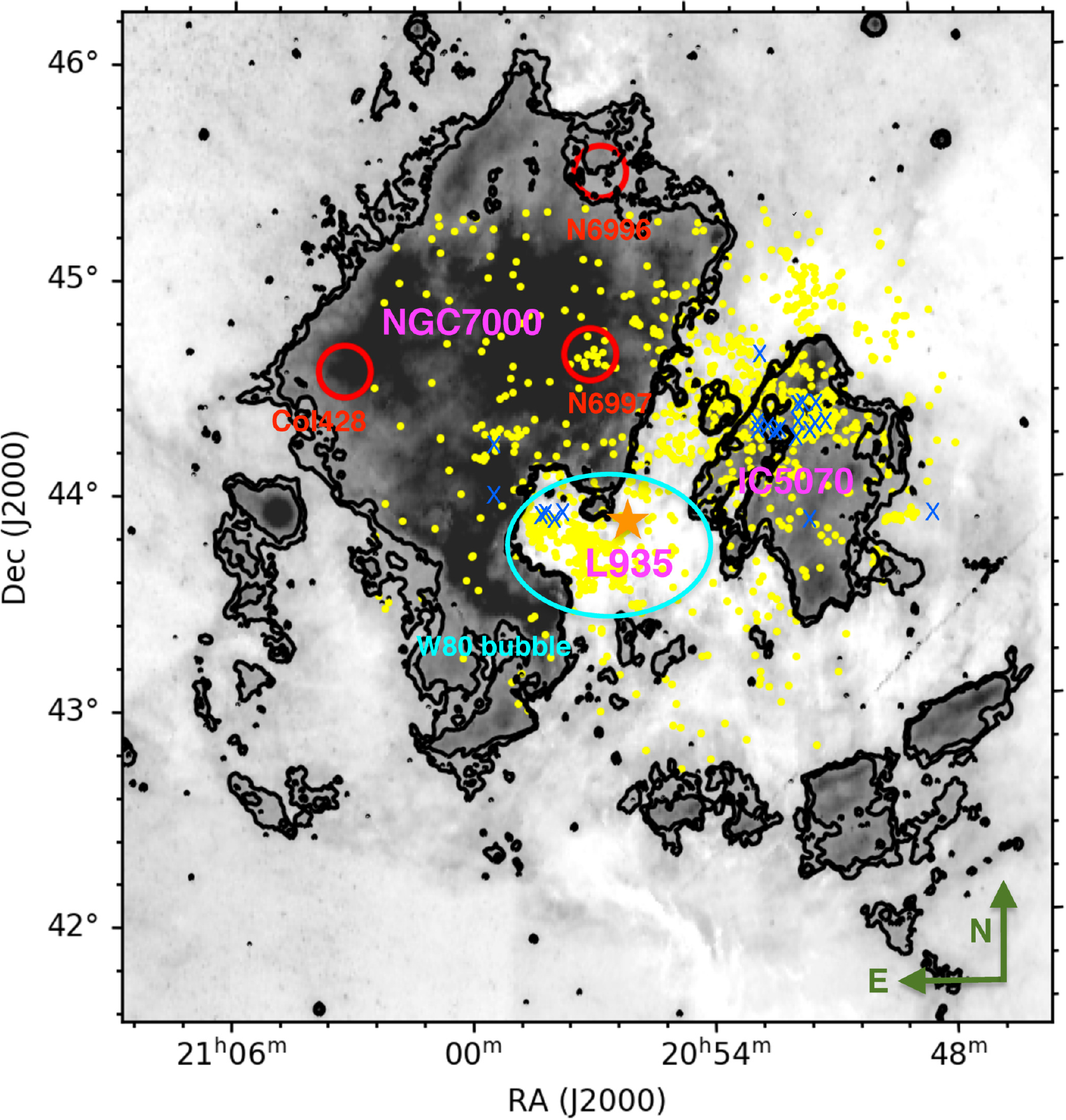}
		\caption{The DSS (red band) image of the W80 region covering the NGC 7000 (North America Nebula), the IC 5070 (Pelican Nebula), the dark cloud L935, and the W80 bubble (the cyan ellipse). Blue crosses show the positions of T-Tauri type stars. The orange star  shows the position of the O-type star (2MASS J20555125+4352246). The yellow dots are the YSOs identified by \cite{2009apj...697..787g}. The red circles display the positions of three stellar clusters. The black lines show the outer contours of bright nebulae.}
  \label{Fig:w80}
  \end{figure}

Figure~\ref{Fig:w80} shows the DSS optical image of the NAP nebulae, with the positions of the ionizing source, YSOs, and T-Tauri stars associations marked. The positions of three stellar clusters, N6996, N6997, and Col428 \citep{1931AnLun...2....1C} are also shown. The outer contours of the bright nebulae are superimposed on the optical image for a better view.

For the large-scale W80 region with low nebular brightness, only a few optical spectral observations have been made. Even in these observations, the amount of data are very limited.
\cite{1973MNRAS.165..381P} have obtained the H${\alpha}$ emission line observations of the W80 complex, and  presented the average radial velocities around -16 km s$^{-1}$ of the IC 5070, NGC 7000, and W80 complex firstly.
\cite{1983apj...269..164f} have made the spectral observations of H$\alpha$ emission lines in 1983 to about 100 positions of the W80 region with Multi-slit Echelon Spectrograph.
\cite{2003apjs..149..405h}  have observed only five positions with the Wisconsin H-alpha Mapper (WHAM) in 2003. The most recent spectral observations are made by \cite{2013RAA....13..547A}, and 26 spectra obtained with employed the Dual Etalon Fabry Perot Optical Spectrometer (DEFPOS). The data mentioned above are contributed greatly to our understanding of the physical conditions in the W80 region. 

Nowadays, the LAMOST Medium-Resolution Spectroscopic Survey of Nebulae (MRS-N) project \citep{2004ChJAA...4....1S,2012raa....12.1197c,2012RAA....12..723Z,2021RAA....21...96W} provides us with a great opportunity to make spectral observations of the W80 Region. With multi-fibers and the 5 degree wide FoV of LAMOST, almost 4000 spectra can be obtained by one exposure. 
The closest spacing between optical fibers is about 2 arcmin. In this study of the W80 Region, the 20 square degree field of view covers not only the W80 complex but also the western region of the bright nebula LBN 391 (LBN 087.23-03.80). LBN 391 is also known as Sharpless 2-119 Region (SH 2-119) \citep{1976A&AS...25...25D}. In this work, we refer to this region as the LBN 391 nebula. Based on the large spatial coverage of LAMOST, we perform an almost complete spectral investigation for the W80 Region.

The electrons in nebular gas are excited or ionized by the ultraviolet radiation of hot stars. The recombination or de-excitation of electrons produces H${\alpha}$ line ($\sim$ 6563 \AA ) emission. The forbidden lines are usually detected in a low-temperature and low-density astrophysical environment.
The [\ion{N}{ii}] at 6548 and 6584 \AA, [\ion{S}{ii}] at 6716 and 6731 \AA, [\ion{O}{ii}] at 3727 \AA, and [\ion{O}{iii}] at 4959 and 5007\AA\,are several common forbidden lines \citep{1952PhRv...87.1002P}.
By optical spectra, different emission lines  can provide a very useful tool to identify the contributions of different radiation mechanisms such as photoionization and shock excitation. 

This paper is organized as follows. The observational data reduction of the W80 Region is described in Section 2.
The observational results, including the relative intensity, the radial velocities (RVs), the FWHM, and the electron density are presented in Section 3. In Section 4, some interesting features are discussed. Section 5 gives a summary.

\section{Data reduction}
\label{sect:obs}	
\subsection{Data Selection}
The focal plane of the LAMOST is circular, and has 4000 fibers evenly equipped within a diameter of 5$^\circ$, and each fiber can move within a radius of 2 arcminutes. The unique design enables LAMOST to obtain 4000 spectra by one exposure. By 2017, LAMOST has completed a five-year regular sky survey with low spectral resolution (R $\sim$ 1800). 
In October 2018, the MRS-N has been officially initiated with higher spectral resolution (R $\sim$ 7500, \cite{2021RAA....21...96W}). The MRS-N survey includes both blue band (from 4950 \AA\ to 5350 \AA) and red band (from 6300 \AA\ to 6800 \AA). 

The MRS-N covers about 1,700 square degrees in the range of 40$^\circ < l < 215^\circ$ and -5$^\circ < b < 5^\circ$ along the Galactic plane. Some bright stars and a large number of Galactic nebulae are set to be observed. Detailed information about the MRS-N can refer to \cite{2021RAA....21...96W,2021RAA....21...51R,2021RAA....21..280Z}, and \cite{wu2021data}. \cite{2021RAA....21...96W} have presented the survey plan of LAMOST-MRS-N and the detailed scientific goals.  \cite{2021RAA....21...51R} have shown the radial velocity calibrations toward the LAMOST medium-resolution spectroscopic survey of nebulae. \cite{2021RAA....21..280Z} have provided the method for eliminating geocoronal H$\alpha$ emissions from the LAMOST-MRS-N spectra. \cite{wu2021data}
have described in detail the MRS-N pipeline with each data processing phase.  In this work, we have made data reduction following the same pipeline.

The W80 Region was observed on November 08, 2020, with the exposure time of 3 $\times$ 900 sec. After the standard data reduction for the MRS-N spectra using the pipeline, the invalid data with a signal-to-noise ratio (S/N) of 999 were deleted first, and about 3800 spectra were obtained. In order to ensure that all the spectra are nebular, the stellar spectra were rejected. 
Then Gaussian fitting was employed to all the emission lines, and only the spectra with high S/N ratios (S/N $>$ 10) for all the H$\alpha$, [\ion{N}{ii}], and [\ion{S}{ii}] emission lines were selected for reliable analysis accuracy. We finally obtained a total of 2982 spectra for the W80 Region.

\begin{figure}[!htp]
  \centering
  \includegraphics[width=0.5\textwidth,angle=270]{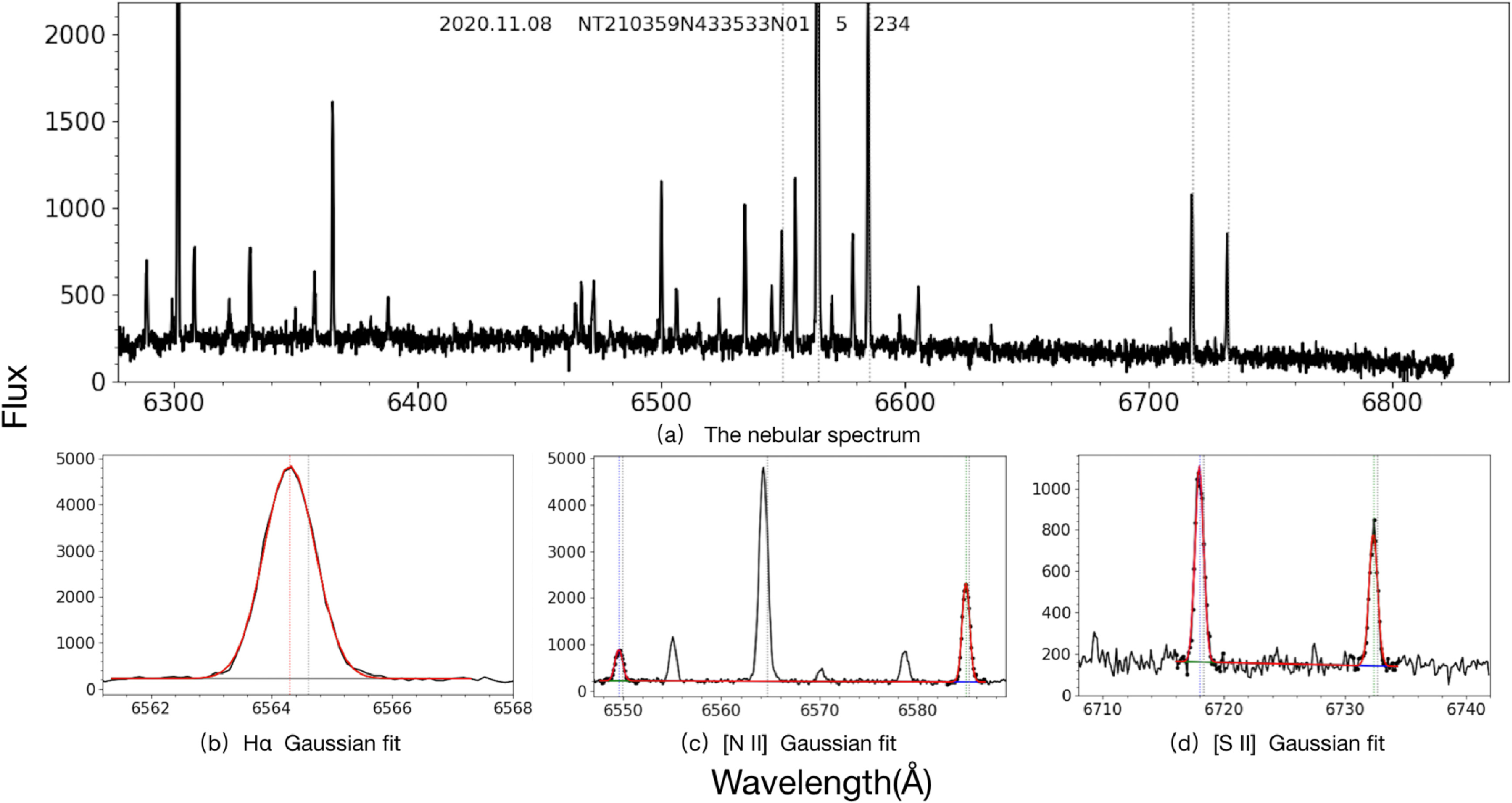}
	  \caption{ One example of the selected 2982 nebular spectra toward the W80 region, showing with the Gaussian fitting results of H$\alpha$, [\ion{N}{ii}], and [\ion{S}{ii}] emission lines. } 
 \label{Fig:fit}
\end{figure}

Figure~\ref{Fig:fit} presents one  typical spectrum as an example of the 2982 spectra, together with the Gaussian fitting (the smallest variance $\chi^{2}$) for the H$\alpha$, [\ion{N}{ii}], and [\ion{S}{ii}] emission lines. The spectral fitting provides us with the measurements of the RVs, the FWHMs, and the relative intensities of the H${\alpha}$, [\ion{N} {ii}], and [\ion{S} {ii}] emission lines, respectively. For the H$\alpha$ line with a S/N ratio of 146, the Gaussian fitting gives a RV uncertainty of 0.1 km s $^{-1}$ and FWHM$_\mathrm{obs}$ uncertainty of 0.4 km s $^{-1}$,
and for the [\ion{N}{ii}] line with a S/N ratio of 89 , RV uncertainty of 0.2 km s $^{-1}$ and FWHM$_\mathrm{obs}$ uncertainty of 0.5 km s $^{-1}$;  
the [\ion{S}{ii}] line has a relatively poor S/N ratio of 71, and its RV uncertainty is 0.3 km s $^{-1}$ and FWHM$_\mathrm{obs}$ uncertainty of 0.8 km s $^{-1}$. There was no further flux calibration made, and we simply employed for analysis of the intensities relative to those of the $\lambda$6554 skyline.

\begin{figure}[!htp]
  \centering
  \includegraphics[width=0.7\textwidth,angle=0]{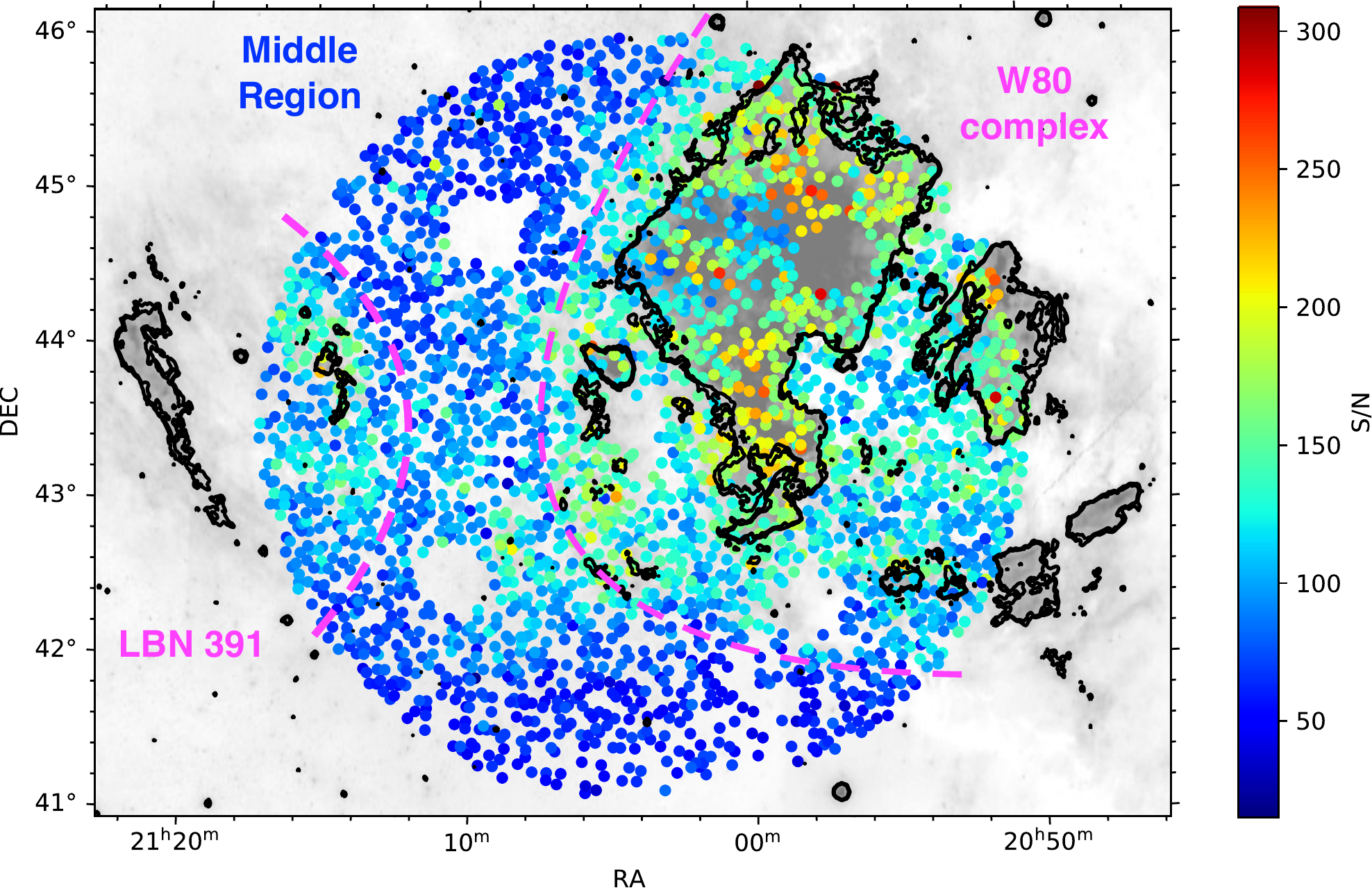}
	  \caption{The spatial distribution of the selected 2982 nebular spectra of the H${\alpha}$ emission lines toward the W80 region, with point colors representing the S/N ratios(S/N $>$ 10). The color bar shows the range of S/N ratios from 10 to 320. The four blank circles are the guider positions on the LAMOST focal plane and lack of data. The two magenta dashed lines divide the LBN 391 nebula region, the W80 complex, and the Middle Region. }
  \label{Fig:snr}
  \end{figure}

Figure~\ref{Fig:snr} presents the spatial distribution of the selected 2982 nebular spectra of H$\alpha$ emission lines, with all the points satisfying S/N $>$ 10. We note that the spectral data points are more evenly distributed over the W80 Region, and the data are much more than the previous spectral observations. The areas of bright nebulae NGC 7000, IC 5070 and LBN 391 have higher S/N ratios. The two magenta dashed lines divide the LBN 391 nebula region, the W80 complex, and the Middle Region.

\begin{figure}[!htp]
  \centering
  \includegraphics[width=0.8\textwidth]{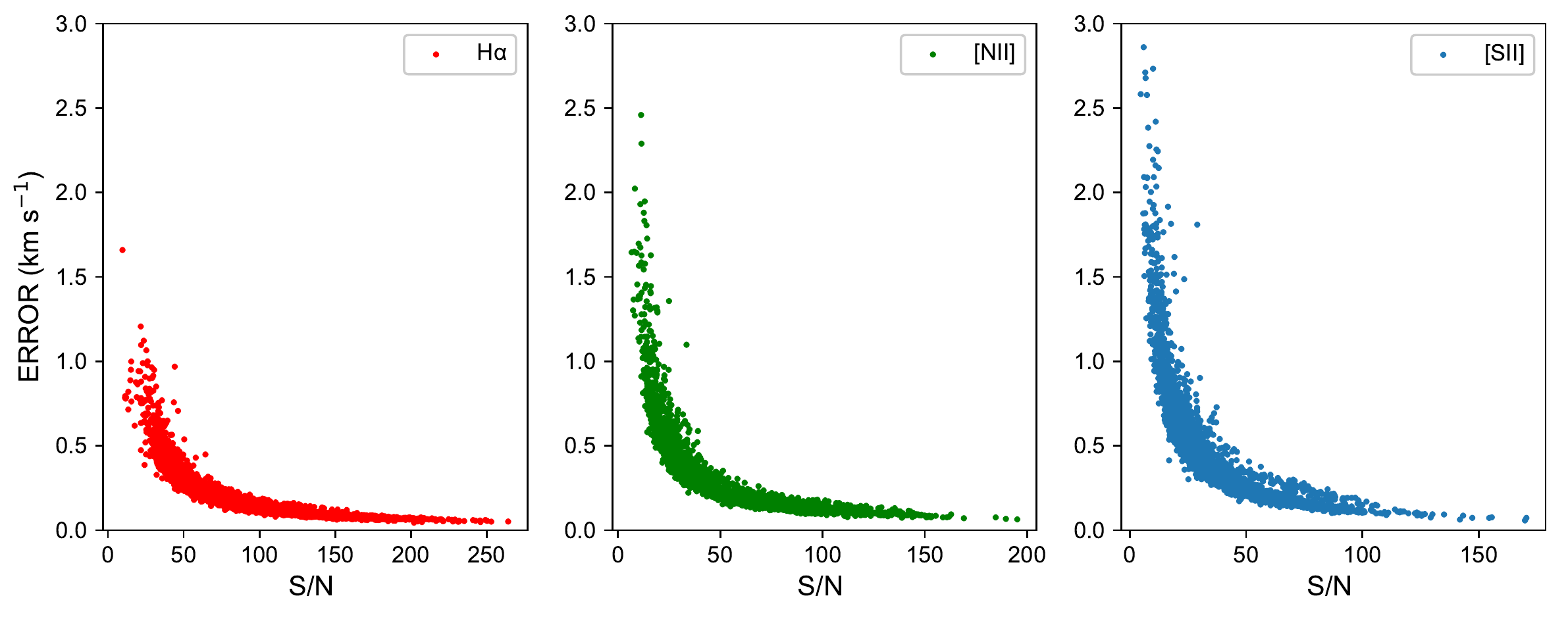}
		  \caption{ The relation of RV uncertainties of H$\alpha$, [\ion{N}{ii}], [\ion{S}{ii}] emission lines with the S/N ratios for the selected 2982 spectra. 
		  } 
 \label{Fig:error}
\end{figure}

Figure \ref{Fig:error}  presents the RV uncertainties of the three emission lines with the S/N ratios, for the selected 2982 spectra. 
There are 99.8$\%$ RV uncertainties less than 1.0 km s $^{-1}$ for the H$\alpha$ lines, 99.3$\%$ RV uncertainties less than 1.5 km s $^{-1}$ for the [\ion{N}{ii}] lines, and for the [\ion{S}{ii}] lines, 99.3$\%$ RV uncertainties less than 2.0 km s $^{-1}$. 
It is reasonable for data analysis to take the RV uncertainties less than 2 km s $^{-1}$ for the three emission lines.  

\subsection{Calibration of FWHM}
The broadening of emission lines is generally caused by a natural broadening of particles, elastic collisions between particles, energy disturbances, and thermal motion of radiating particles. 
Except for these physical broadening, instrumental broadening is another important factor. We need to subtract the instrumental broadening effect from the observed FWHMs. 

For spectral observations, the standard arc lines and skylines are commonly used to estimate the instrumental broadening. \cite{2021RAA....21..280Z} have compared one by one the line dispersion of the arc lines with the $\lambda$6554 skyline of LAMOST MRS-N spectra, and found that the line dispersion of $\lambda$6554 is identical with that of the arc line. Because the values of line dispersion are different for different fibers, in this work, the instrumental broadening of each spectrum has been removed with its own $\lambda$6554 skyline.

The FWHM correction has been made according to equation \ref{eq:gaizheng} \citep{1977ApJ...211..115R}, and the FWHM uncertainties are calculated by using the error propagation equation.

\begin{equation}
  \mathrm{FWHM}_{\mathrm{obs}}^{2} = \mathrm{FWHM}_{\mathrm{real}}^{2}  + \mathrm{FWHM}_{\mathrm{inst}}^{2}
	\label{eq:gaizheng}
\end{equation}
where FWHM$_{\mathrm{obs}}$ is the observed FWHMs, $\mathrm{FWHM}_{\mathrm{inst}}$ is the instrumental FWHMs, and $\mathrm{FWHM}_{\mathrm{real}}$ is the corrected FWHMs, the unit of FWHMs are km s$^{-1}$. With no more specific notes, all the FWHMs in this work represent the corrected FWHMs ($\mathrm{FWHM}_{\mathrm{real}}$, the $\sigma_{\mathrm{FWHM}_{\mathrm{obs}}}$ is the uncertainties of $\mathrm{FWHM}_{\mathrm{obs}}$ and the $\sigma_{\mathrm{FWHM}_{\mathrm{inst}}}$ is the uncertainties of $\mathrm{FWHM}_{\mathrm{inst}}$). 
Because the $\lambda$6554 skyline is a little far from the [\ion{S} {ii}] $\lambda\lambda$ 6716, 6731 \AA \  lines, which could induce larger uncertainties, we present the FWHM results and related analysis only for the H${\alpha}$ and [\ion{N} {ii}] emission lines.

Further criteria have been set to control the analysis accuracy to avoid  contamination by the spectra with larger FWHM uncertainties. First, we selected the spectra that the $\mathrm{FWHM}_{\mathrm{obs}}$ were greater than the FWHM$_{\mathrm{inst}}$,   
that is, $\mathrm{FWHM}_{\mathrm{obs}}$- $\sigma_{\mathrm{FWHM}_{\mathrm{obs}}}$ $\textgreater$  $\mathrm{FWHM}_{\mathrm{inst}}$ + $\sigma_{\mathrm{FWHM}_{\mathrm{inst}}}$,  the $\sigma_{\mathrm{FWHM}_{\mathrm{obs}}}$ is the uncertainties of $\mathrm{FWHM}_{\mathrm{obs}}$,  95.7$\%$ $\sigma_{\mathrm{FWHM}_{\mathrm{obs}}}$ of H$\alpha$ lines are less than 1.5 km s $^{-1}$, and 95.4$\%$ $\sigma_{\mathrm{FWHM}_{\mathrm{obs}}}$ of [\ion{N}{ii}] lines are less than 2.5 km s $^{-1}$; and the $\sigma_{\mathrm{FWHM}_{\mathrm{inst}}}$ is the uncertainties of $\mathrm{FWHM}_{\mathrm{inst}}$,  96.6$\%$ $\sigma_{\mathrm{FWHM}_{\mathrm{inst}}}$ are less than 2 km s $^{-1}$. And then, we selected the spectra satisfying 
$\mathrm{FWHM}_{\mathrm{real}}$ \textgreater 3 $\sigma_{\mathrm{FWHM}_{\mathrm{real}}}$, the $\sigma_{\mathrm{FWHM}_{\mathrm{real}}}$ is the uncertainties of $\mathrm{FWHM}_{\mathrm{real}}$. 
We finally obtained 1903 spectra for FWHM analysis in the whole region, for both the H$\alpha$ and [\ion{N}{ii}] emission lines.
For the 1903 spectra, we find that 
99.1$\%$ FWHM uncertainties of H$\alpha$ lines are less than 4 km s $^{-1}$, and 94.3$\%$ FWHM uncertainties of [\ion{N}{ii}] lines are less than 5 km s $^{-1}$. 
It is reasonable for us to take the FWHM uncertainties of 5 km s $^{-1}$ for further analysis. 

\section{result}
\label{sect:data}
The LAMOST MRS-N project provides us with an effective method to understand the whole nebula region in detail. As shown in Figure \ref{Fig:snr}, the 2982 spectra we obtained for the H${\alpha}$, [\ion{N} {ii}], and [\ion{S} {ii}] emission lines all have high S/N ratios and reasonable spatial distribution, confirming the existence of emission lines in the whole region.

The relative intensities, radial velocities, and FWHMs of the H${\alpha}$, [\ion{N} {ii}], and [\ion{S} {ii}] emission lines are measured for the W80 Region. To make the presentation clear, the following sections begin with the spatial distribution, and follow by the histogram analysis of the W80 complex. The electron density is calculated with the intensity ratios of double [\ion{S} {ii}] emission lines.

\subsection{The relative intensity}
The intensities discussed in this work are the relative intensities corrected by the simultaneously observed $\lambda$6554 skylines of each fiber, for all the H${\alpha}$, [\ion{N} {ii}], and [\ion{S} {ii}] emission lines.

\begin{figure}[!htp]
\begin{minipage}{0.48\linewidth}
	\centerline{\includegraphics[width=7.5cm]{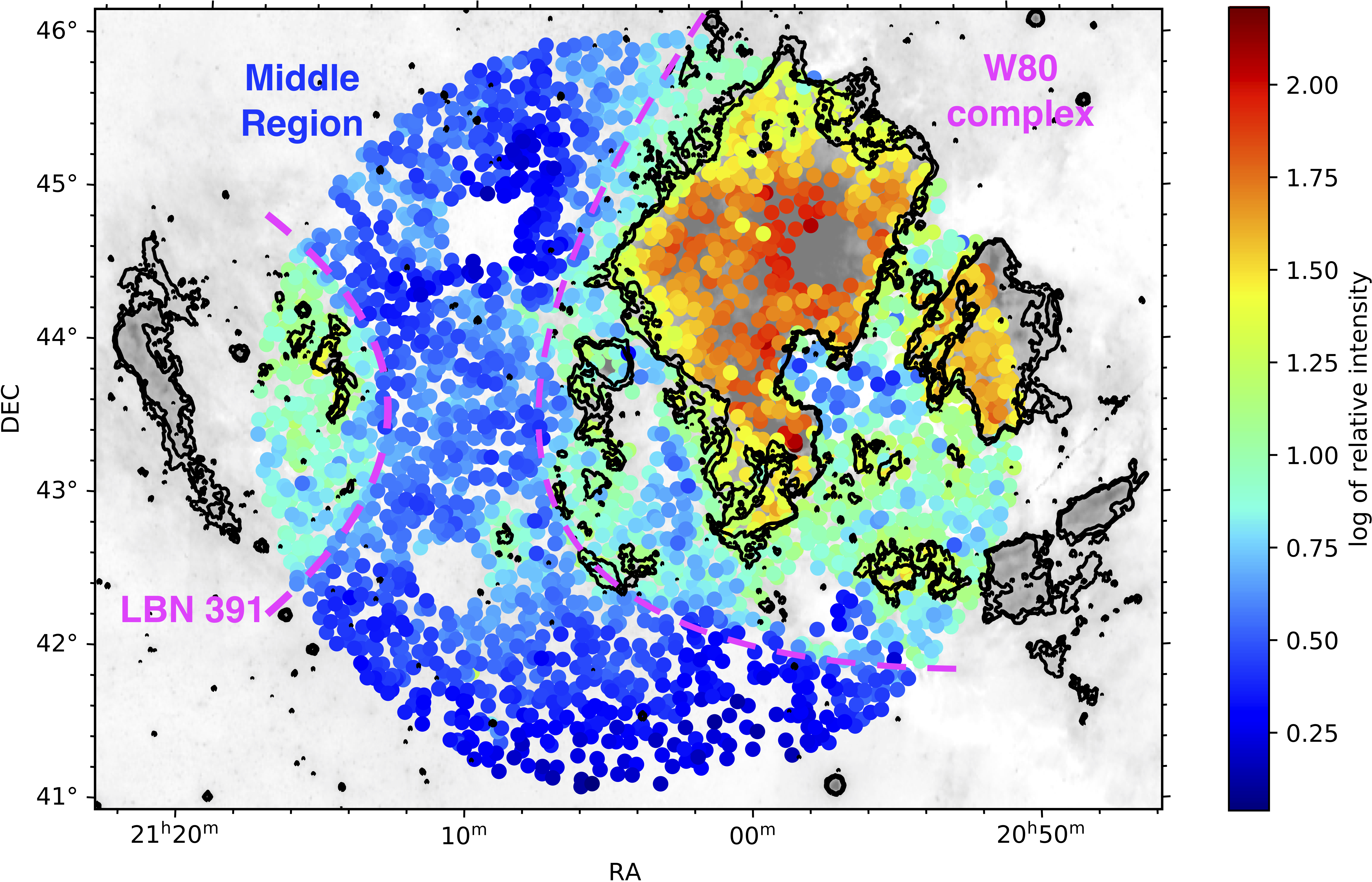}}
  \centerline{(a) H${\alpha}$  }
\end{minipage}
\hfill
\begin{minipage}{0.48\linewidth}
  \centerline{\includegraphics[width=7.5cm]{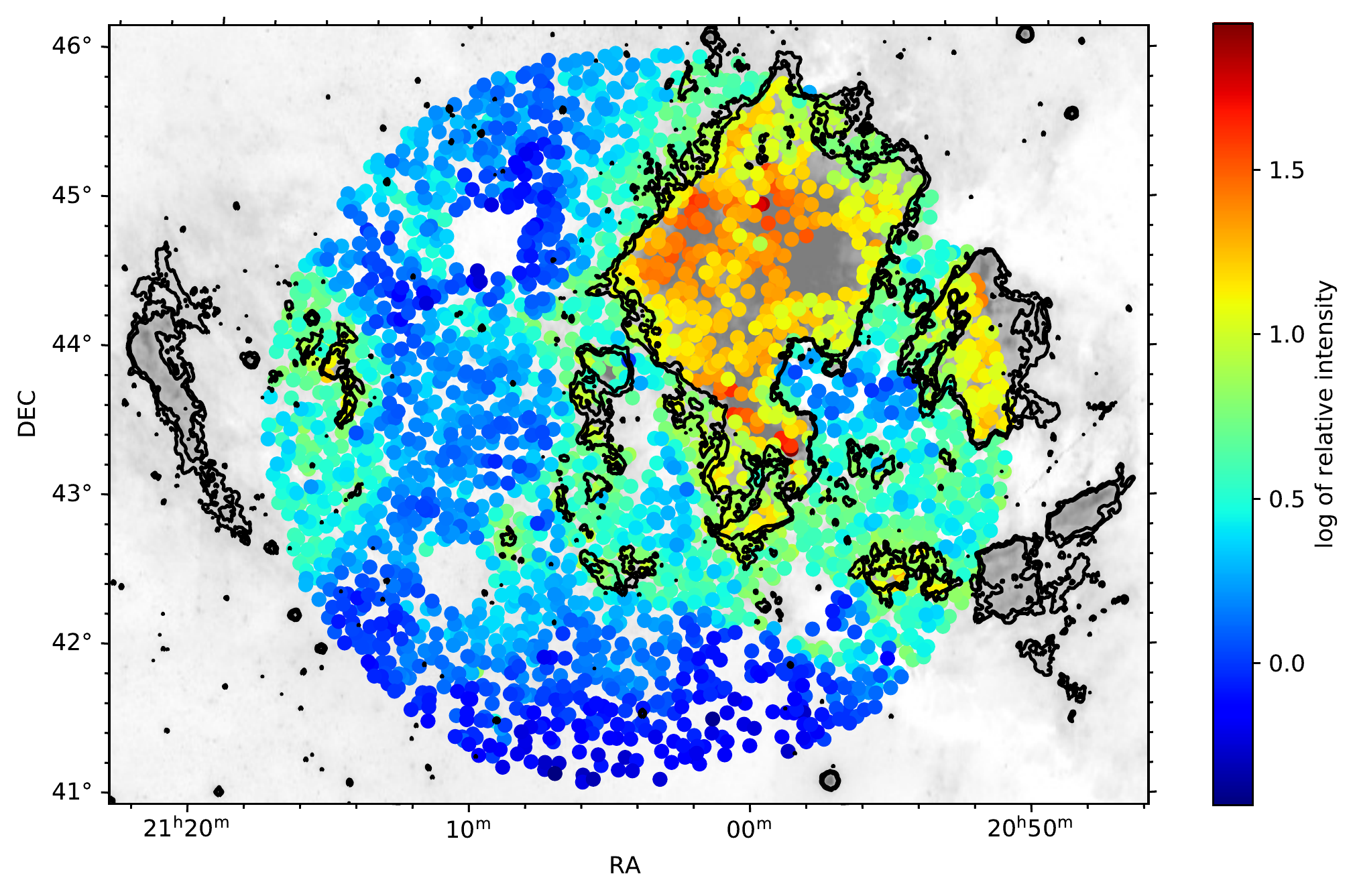}}
	\centerline{(b) [\ion{N} {ii}] }
\end{minipage}
\vfill
\begin{minipage}{0.48\linewidth}
  \centerline{\includegraphics[width=7.5cm]{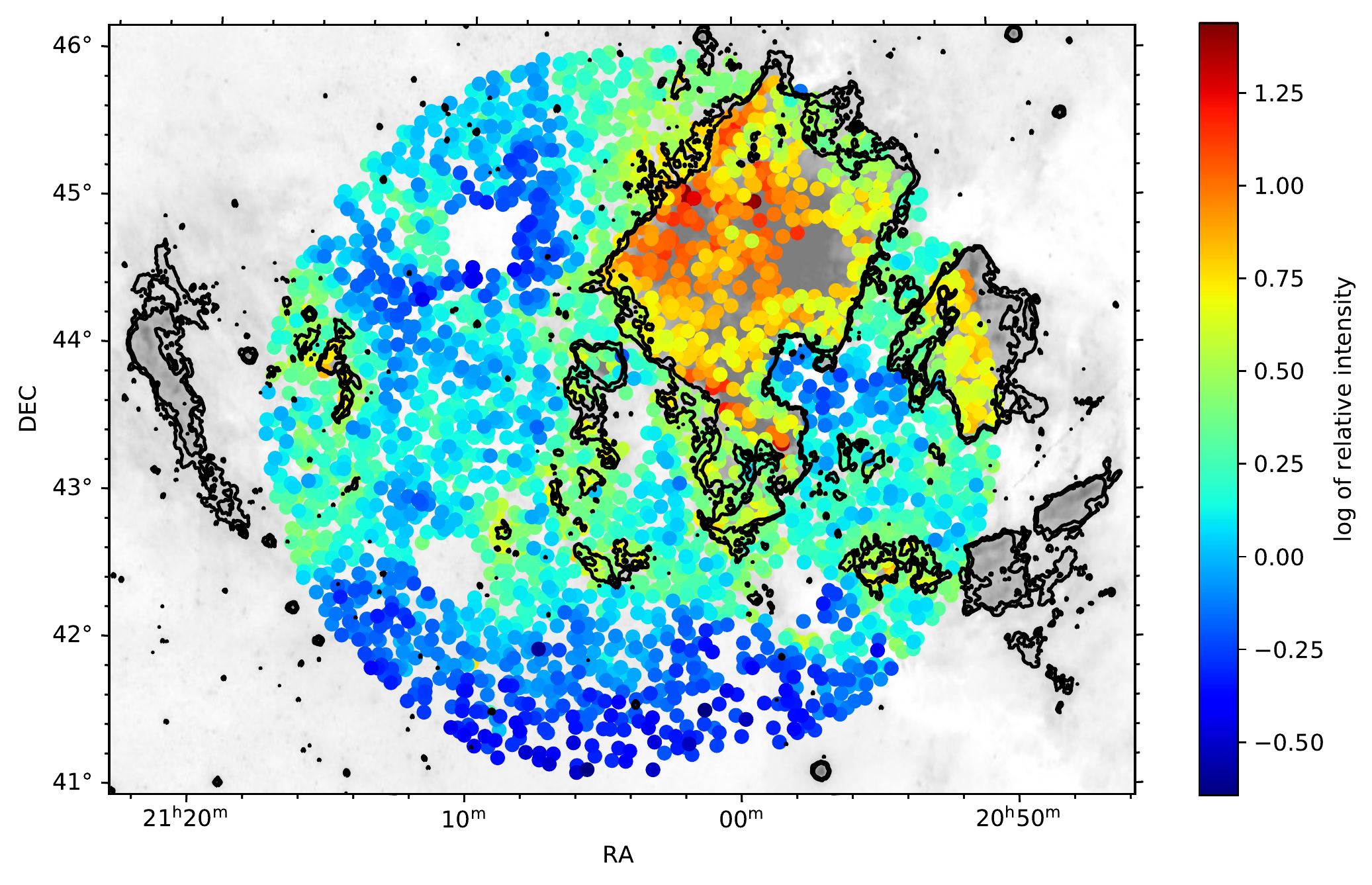}}
  \centerline{(c)  [\ion{S} {ii}]}
\end{minipage}
\caption{ The spatial distributions of the relative intensities of (a) H${\alpha}$, (b) [\ion{N} {ii}] and (c) [\ion{S} {ii}] emission lines. 
The magenta dashed lines in (a) indicate the LBN 391 nebula and the W80 complex, and the Middle Region is denominated the region between them.
The color bar shows the log values of relative intensities. }
\label{Fig:RE}
\end{figure}

Figure \ref{Fig:RE} shows the spatial distribution of relative intensities, with colors ranging from blue to red indicating weak to strong intensity. 
The western part of the FoV is the W80 complex, which contains the NGC 7000, L935 and IC 5070 regions. The eastern part of the FoV is the bright nebula LBN 391. The area between the LBN 391 nebula and the W80 complex in this work is called  Middle Region.

It is worth noting that the H${\alpha}$ emission in the NGC 7000 region is mainly concentrated in the central part, while the stronger [\ion{N} {ii}] and [\ion{S} {ii}] emissions spread at the northeastern edge. This feature is particularly evident for the [\ion{S} {ii}] emissions. 

\begin{figure}[!htp]
  \centering
  \includegraphics[width=0.9\textwidth,angle=0]{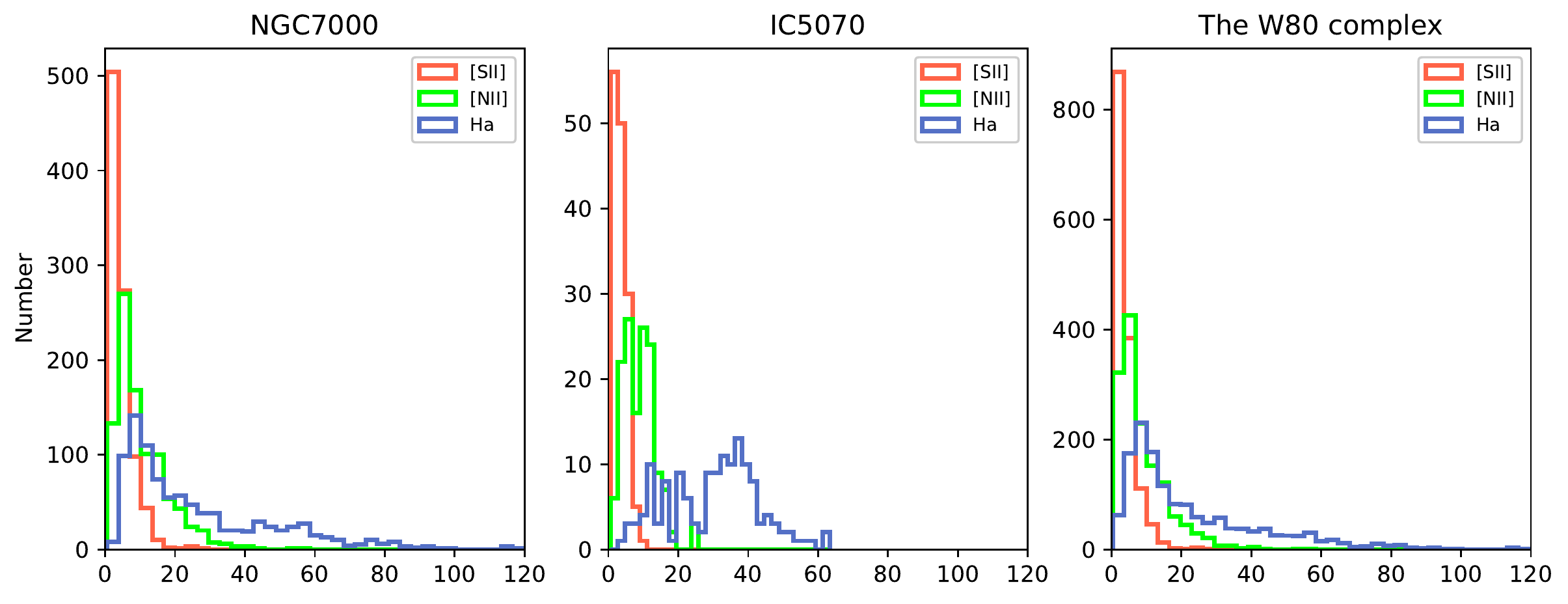}
   \caption{The histograms of relative intensities for NGC 7000, IC 5070, and the W80 complex.}
  \label{Fig:amp_ha}
  \end{figure}

 Figure\ref{Fig:amp_ha} shows histograms of relative intensities for the three regions. The left panel is for NGC 7000, the middle is IC 5070 and the right is the W80 complex. The intensities of IC 5070 are the weakest in the three regions, and the H$\alpha$ emission appears a peak in the range of 30 to 40. The intensities of NGC 7000 and W80 complex are basically consistent, and the maximum intensity values of H${\alpha}$, [\ion{N} {ii}], and [\ion{S} {ii}] emission are 100,40 and 20, respectively.

\begin{figure}[!htp]
  \centering
  \includegraphics[width=0.9\textwidth,angle=0]{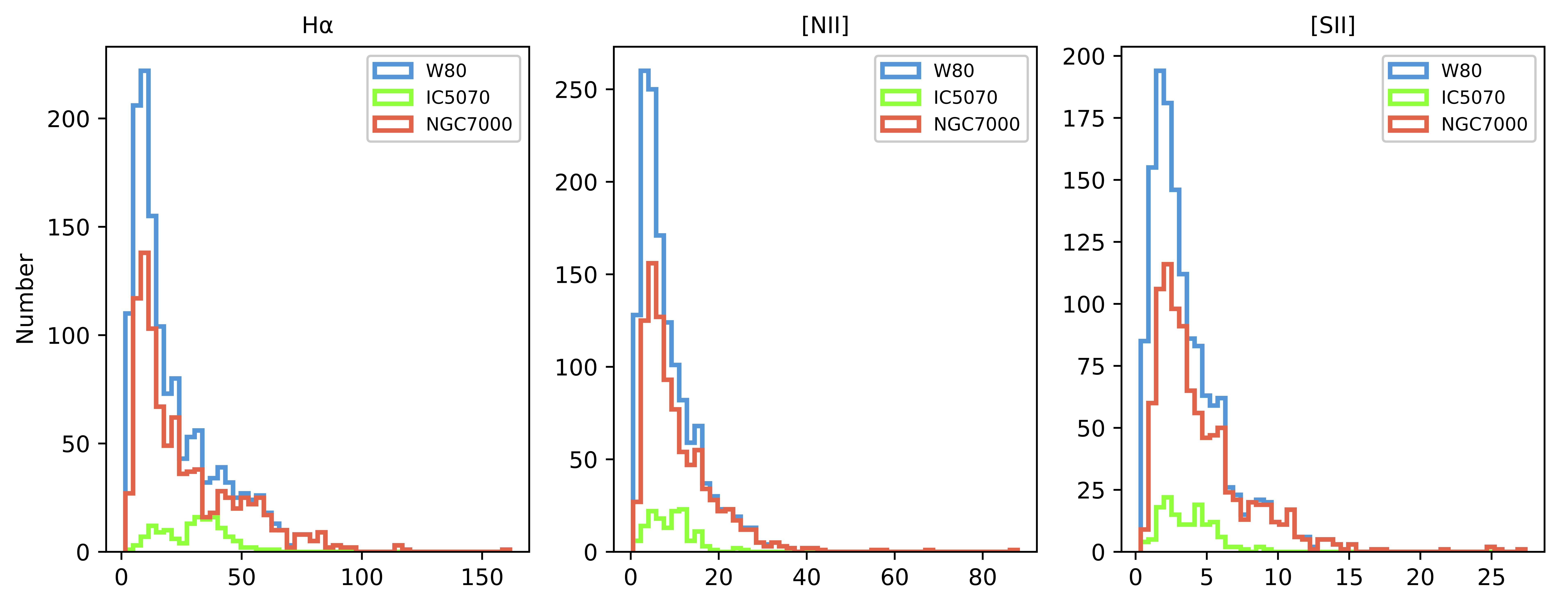}
		\caption{The histograms of relative intensities of the H${\alpha}$, [\ion{N} {ii}], and [\ion{S} {ii}] emission lines. }
  \label{Fig:amp_NGC 7000}
  \end{figure}

Figure \ref{Fig:amp_NGC 7000} presents histograms of relative intensities for three emission lines. The left panel is the intensity distribution of H${\alpha}$, the middle is [\ion{N} {ii}] and the right is [\ion{S} {ii}] emission lines. The H${\alpha}$ emissions have the strongest intensity and [\ion{S} {ii}] emissions are the weakest. 

\subsection{The radial velocity}
All the radial velocities (RVs) presented in this work are corrected to the heliocentric. A positive radial velocity value means the target is moving away from us, while a negative radial velocity means moving towards us.

\begin{figure}[!htp]
\begin{minipage}{0.48\linewidth}
	\centerline{\includegraphics[width=7.5cm]{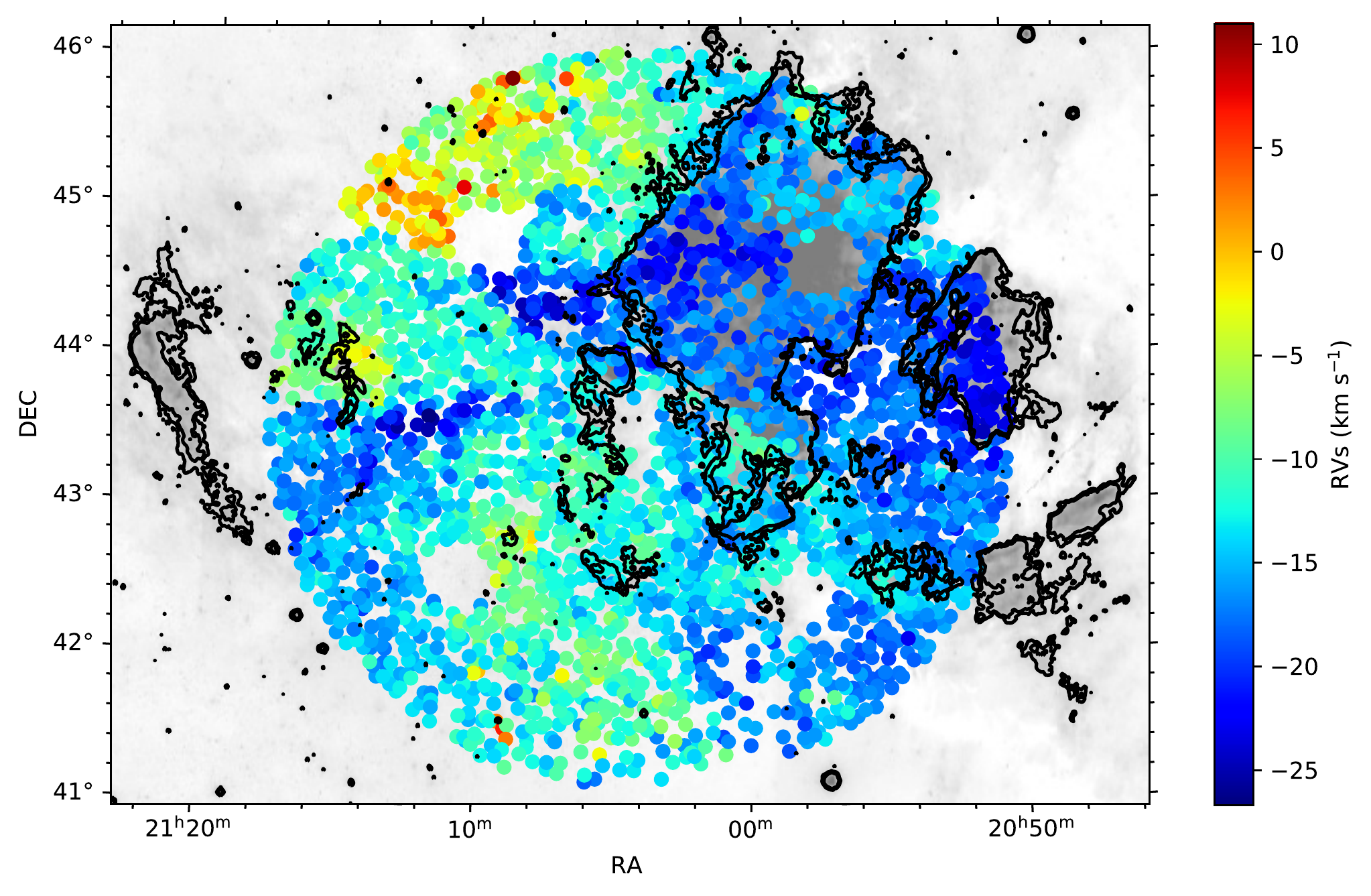}}
  \centerline{(a1) H${\alpha}$  }
\end{minipage}
\hfill
\begin{minipage}{0.48\linewidth}
  \centerline{\includegraphics[width=7.5cm]{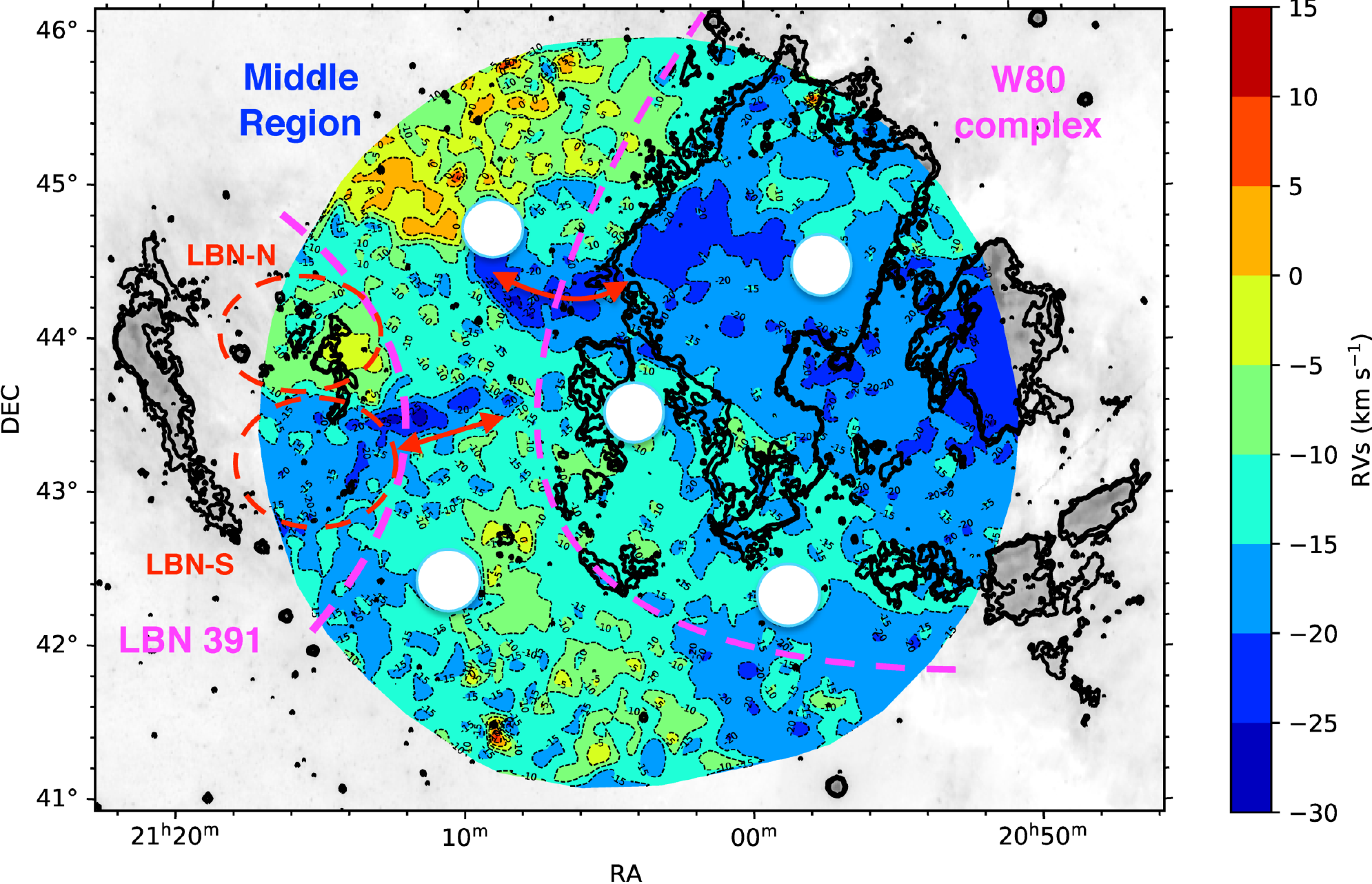}}
  \centerline{(a2) The interpolation map of H${\alpha}$ RVs}
\end{minipage}
\vfill
\begin{minipage}{0.48\linewidth}
  \centerline{\includegraphics[width=7.5cm]{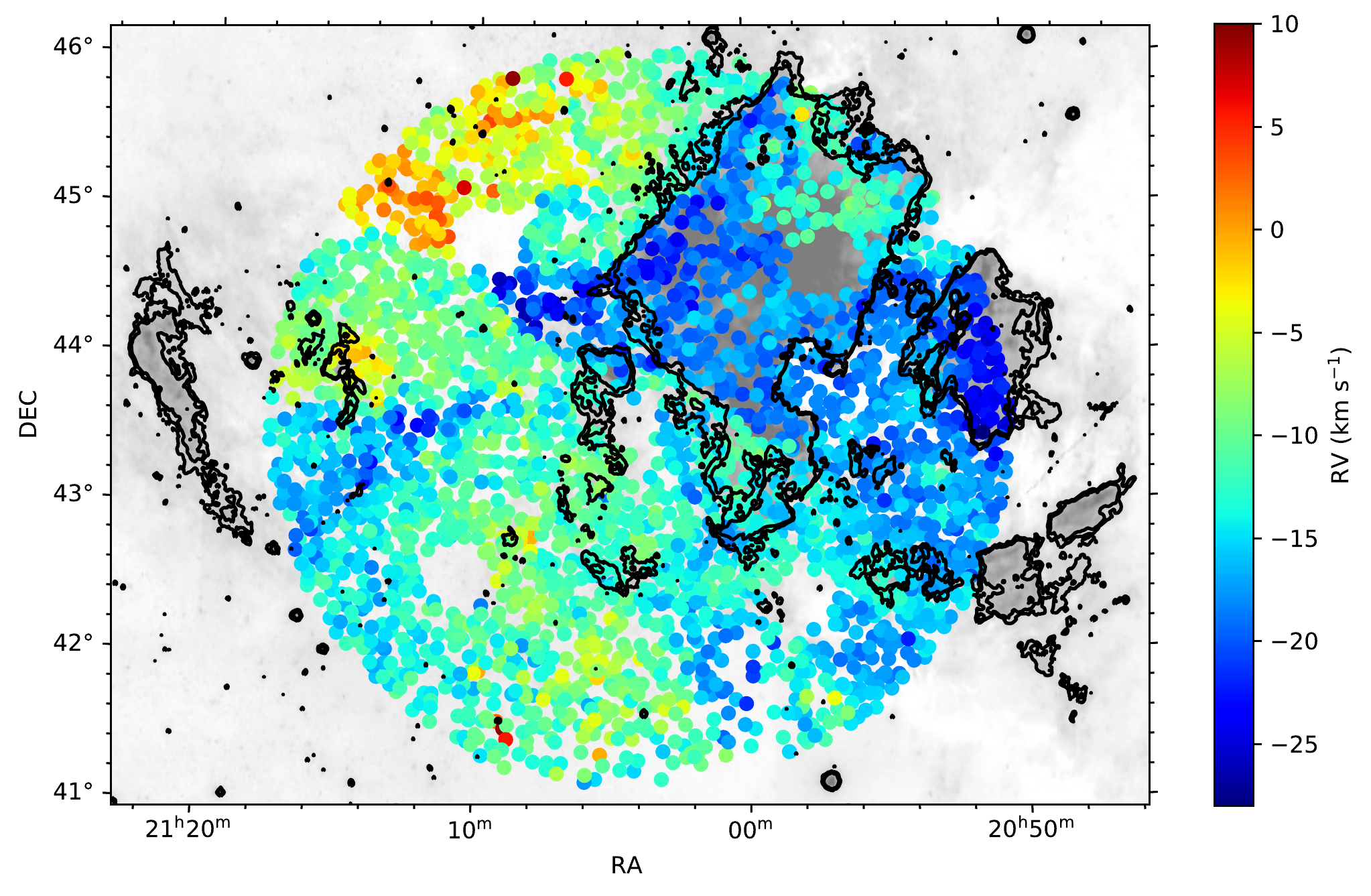}}
	\centerline{(b) [\ion{N} {ii}] }
\end{minipage}
\hfill
\begin{minipage}{0.48\linewidth}
	\centerline{\includegraphics[width=7.5cm]{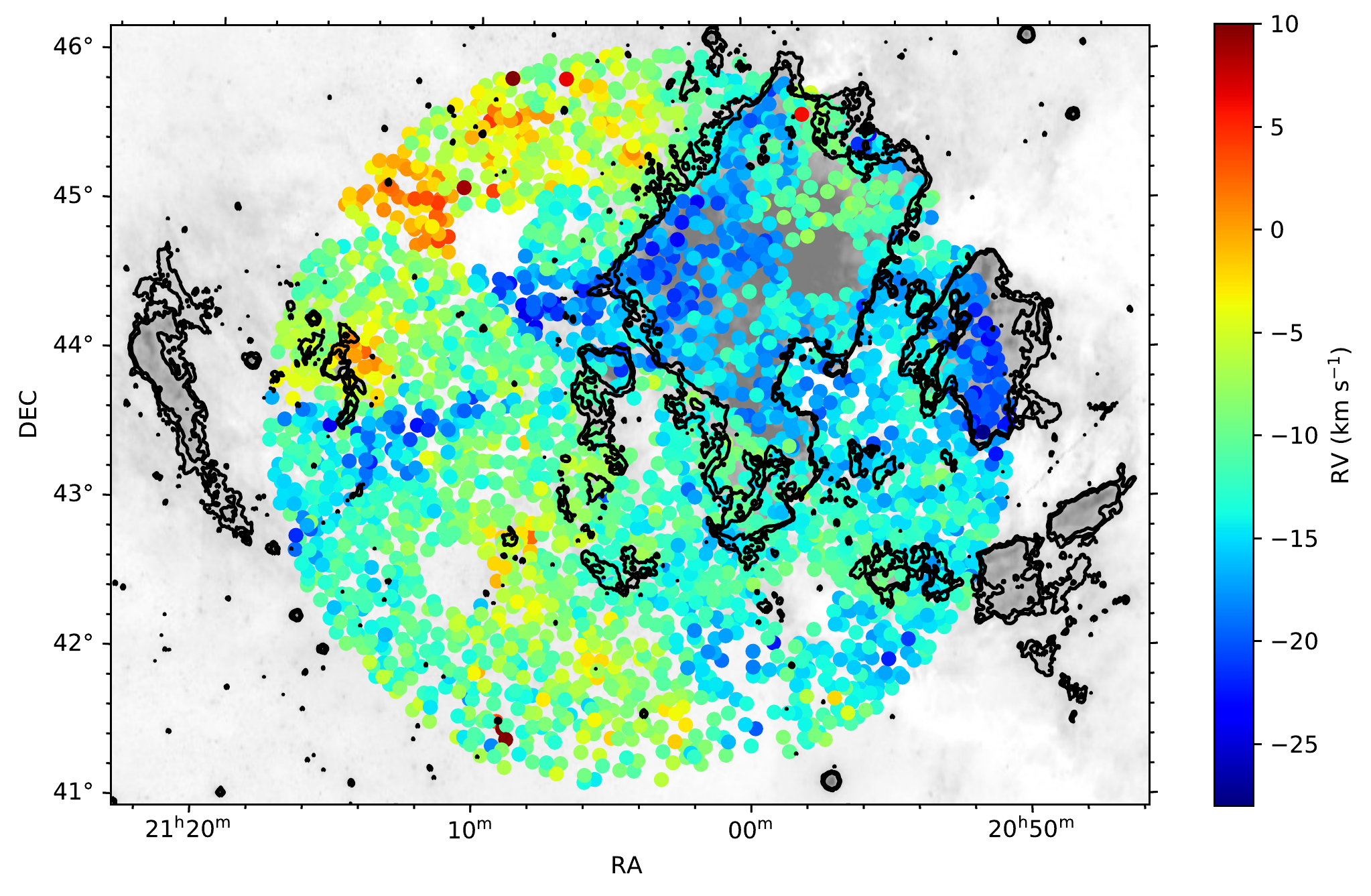}}
  \centerline{(c) [\ion{S} {ii}]}
\end{minipage}
\caption{The spatial distributions of RVs of (a1) H${\alpha}$, (b) [\ion{N} {ii}], and (c) [\ion{S} {ii}] emission lines, with color bars showing the RV values. The interpolation map of H${\alpha}$ RV distribution (a2) is also shown with region notes and feature marks; two red arrow lines indicate a `curved feature' to the east of NGC 7000 and a `jet feature' to the west of LBN 391 nebula, and two red dashed ovals, LBN-N and LBN-S, mark the northern and southern regions of LBN 391 nebula. Five blank circles show the guider positions and so the lack of data. }
\label{Fig:RV}
\end{figure}

The RV spatial distributions of all spectra are presented in Figure~\ref{Fig:RV}. Figure~\ref{Fig:RV} (a1) shows the radial velocity distributions of the H$\alpha$ emission lines. The bright nebulae within W80 complex have a velocity range from -25 to -15 km s$^{-1}$, and the velocity range of -10 to 0 km s$^{-1}$ can be seen in the Middle Region, in which some velocity components of -5 to +5 km s$^{-1}$ are visible in the northern part. The LBN 391 nebula has obviously different velocity regions; its northern region has mainly velocities between -10 to 0 km s$^{-1}$, while the southern region shows velocity values from -25 to -15 km s$^{-1}$, indicating that the two regions must have different velocity components.

Figure\ref{Fig:RV} (a2) further shows the interpolated map of RVs for the H$\alpha$ emission line, in order to reveal the RV features in each region more clearly. 
For the W80 complex, the RVs of the NGC 7000 and IC 5070 are merged together, not being separated as in the intensity map, indicating that the two nebulae are moving towards us at a uniform velocity.
 \citeauthor{1983A&A...124..116W} \citeyearpar{1968ZA.....68..368W,1983A&A...124..116W} have pointed out the NAP are all part of the same extended \ion{H}{ii} region, and \cite{1980ApJ...239..121B} deduced from CO observations that the NAP complex may be a huge molecular cloud being destroyed by early-type stars.

It is worth noting that there is a high-speed `curved feature' extended to the east from the bright nebula NGC 7000, and the RVs of this feature are concentrated at -30 to -25 km s$^{-1}$ (see Figure \ref{Fig:RV}(a2) red arrow line). This feature extends further to the northeast beyond the NGC 7000.
To the LBN 391 nebula, there is also a `jet feature' extended along the east-west direction, 
which is also beyond the intensity scope of the LBN 391 nebula. 
Considering the 2 km s$^{-1}$ uncertainty of the RV measurements and relatively shaped structures, we believe these large RV features are real. The similar RV values of -30 to -25 km s$^{-1}$ can be also seen associated with the bright nebula IC 5070 and dark cloud L935.

\begin{figure}[!htp]
  \centering
  \includegraphics[width=0.9\textwidth,angle=0]{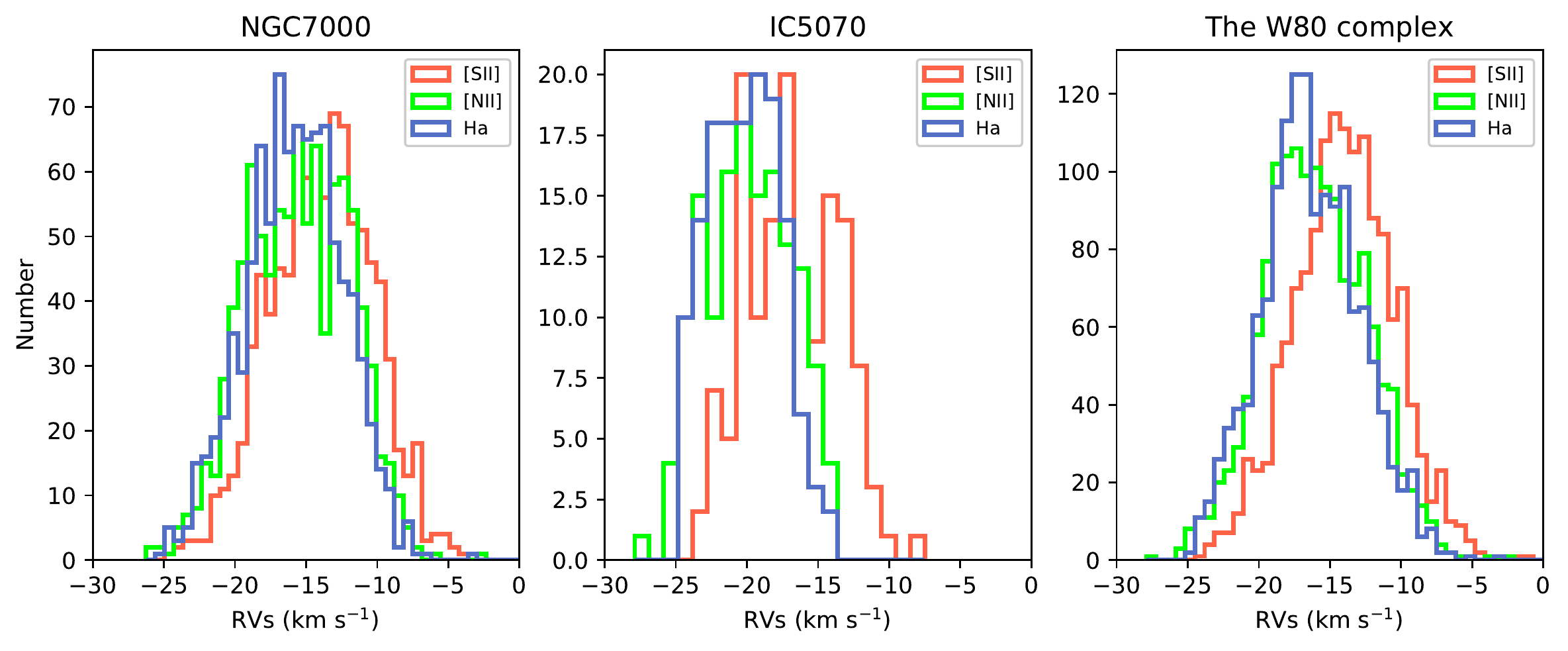}
	  \caption{The histograms of RVs  for NGC 7000, IC 5070, and the W80 region. }
  \label{Fig:rv_ha}
  \end{figure}

In order to study the RV distributions, we present two sets of histograms.
Figure\ref{Fig:rv_ha} shows the histograms for the three regions; the left panel is NGC 7000, the middle is IC 5070 and the right is the W80 complex. Among the three regions, IC 5070 has the largest RV values; the peak RV of H${\alpha}$ emission lines is at -20.1 km s$^{-1}$, the [\ion{N} {ii}] emission lines at -19.9 km s$^{-1}$, and the [\ion{S} {ii}] emission lines at -16.1 km s$^{-1}$. The range of the RV distribution of NGC 7000 is basically similar to that of the W80 complex, and the RVs of H${\alpha}$ emission lines are concentrated at -16.0 km s$^{-1}$, the [\ion{N} {ii}] emission lines at -16.0 km s$^{-1}$, and the [\ion{S} {ii}] emission lines at -13.9 km s$^{-1}$.
We have noticed the fact that our  field of view (see Figure\ref{Fig:snr}) does not fully cover the entire IC 5070 nebula.

Comparing to the RVs of the H$\alpha$,  [\ion{N} {ii}] emission lines, and the RVs of the [\ion{S} {ii}] emission lines display inconsistencies, and there is about a 5 km s$^{-1}$ difference  between the RV peaks. We have in fact checked the RV uncertainties for those bright nebular regions, in which the RV uncertainties of the [\ion{S} {ii}] emission lines  are all distributed below 0.5 km s$^{-1}$. So the RV inconsistencies are real and may indicate different components detected along the line of sight. \cite{2019ApJ...880...16K} have pointed out that different ionization lines of different species probe different zones of a nebula. The [\ion{S} {ii}] emission lines may trace the very outer edge of the nebula.

\begin{figure}[!htp]
  \centering						
		\includegraphics[width=0.9\textwidth,angle=0]{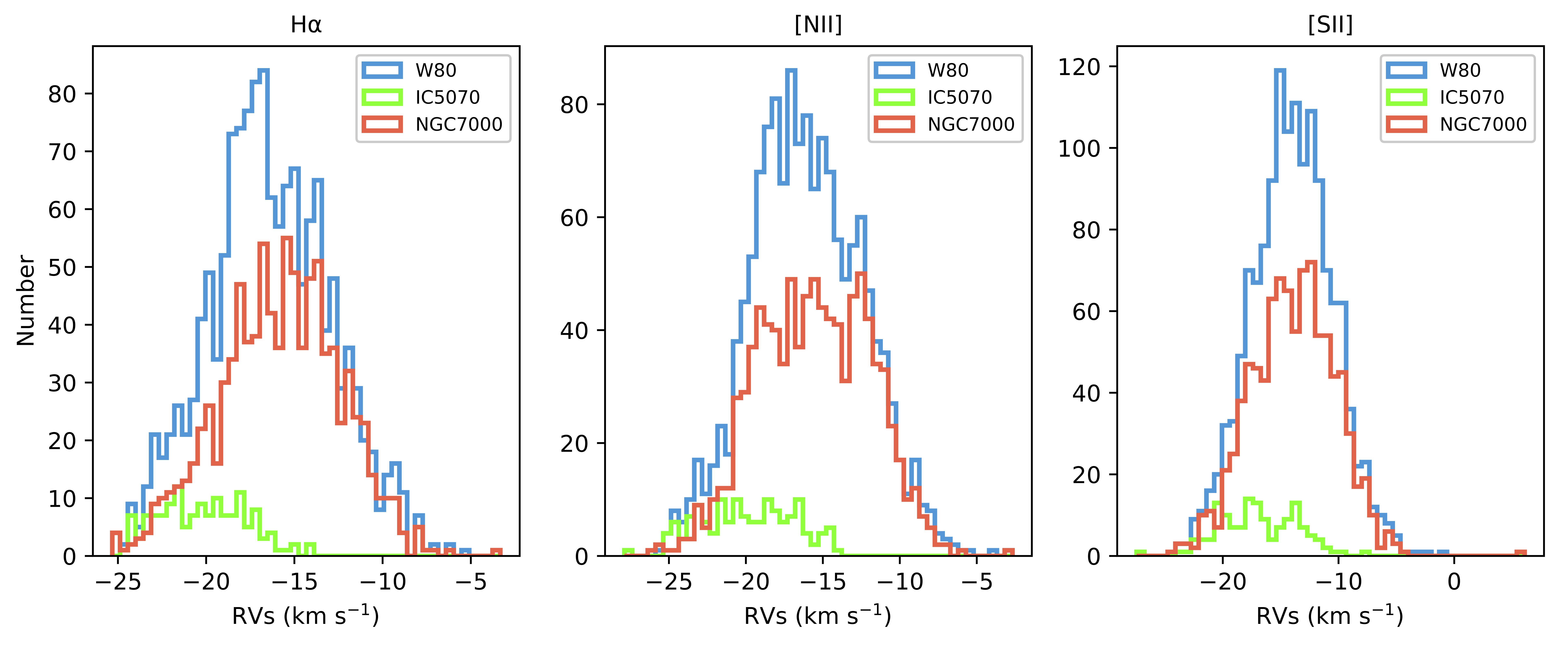}
	  \caption{The histograms of RVs  for the H${\alpha}$,  [\ion{N} {ii}], and [\ion{S} {ii}]  emission lines. }
  \label{Fig:rv_NGC 7000}
  \end{figure}

Figure \ref{Fig:rv_NGC 7000} presents three histograms of RVs for three emission lines; the left panel is H${\alpha}$, the middle is [\ion{N} {ii}] and the right is the [\ion{S} {ii}] emission lines. It can be seen that the [\ion{S} {ii}] emission lines have the lowest RV values in NGC 7000, IC 5070, and W80 complex, and their RVs peak at -13.9, -16.9, and -14.0 km s$^{-1}$. The RV distributions of H$\alpha $ emission lines peak at -15.6, -20.2, and -16.4 km s$^{-1}$, and the [\ion{N} {ii}] emission lines peak at -15.6, -19.9, and -16.0 km s$^{-1}$, respectively. The RV distributions of the H$\alpha $ emission lines and the [\ion{N} {ii}] emission lines are consistent.

\subsection{The FWHM}\label{sec:3.3}
\begin{figure}[!htp]
\begin{minipage}{0.48\linewidth}
  \centerline{\includegraphics[width=7.5cm]{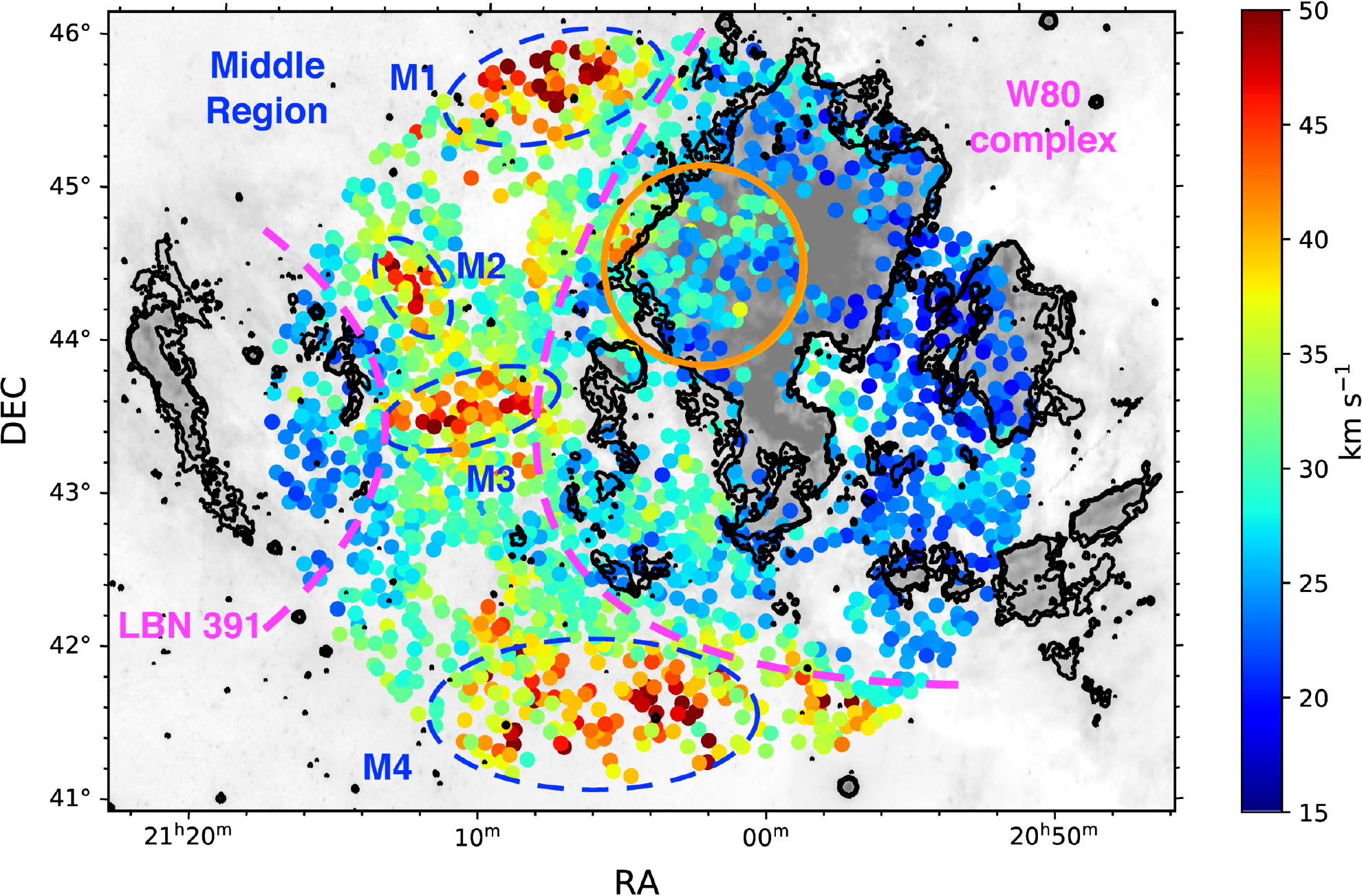}}
  \centerline{(a) FWHMs of H${\alpha}$ emission lines}
\end{minipage}
\hfill
\begin{minipage}{0.48\linewidth}
		\centerline{\includegraphics[width=7.5cm]{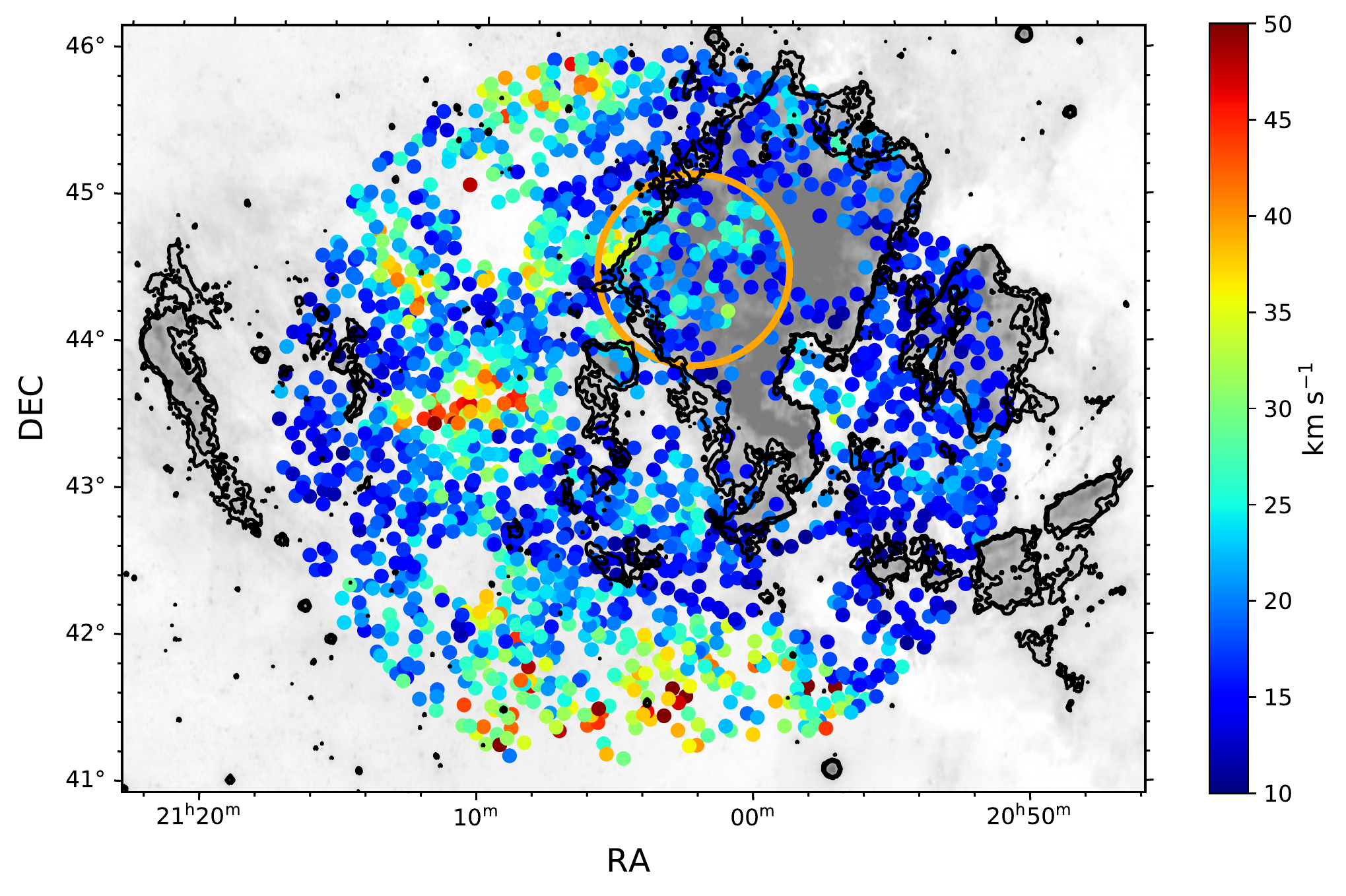}}
	\centerline{ (b)FWHMs of [\ion{N} {ii}] emission lines}
\end{minipage}			
\vfill
\begin{minipage}{0.48\linewidth}
  \centerline{\includegraphics[width=7.5cm]{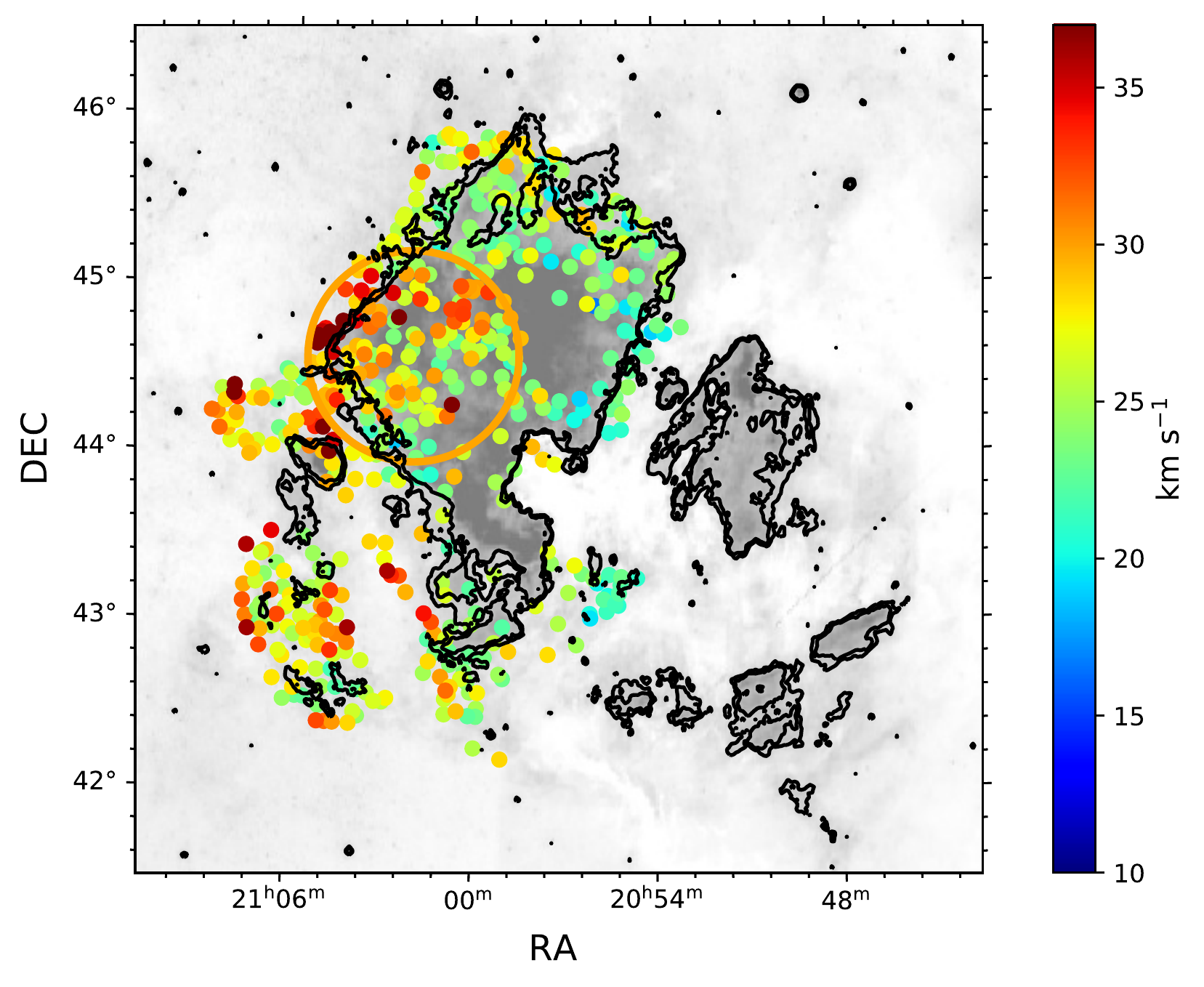}}
  \centerline{(c) FWHMs of H${\alpha}$ emission lines of NGC 7000}
\end{minipage}
\hfill
\begin{minipage}{0.48\linewidth}
		\centerline{\includegraphics[width=7.5cm]{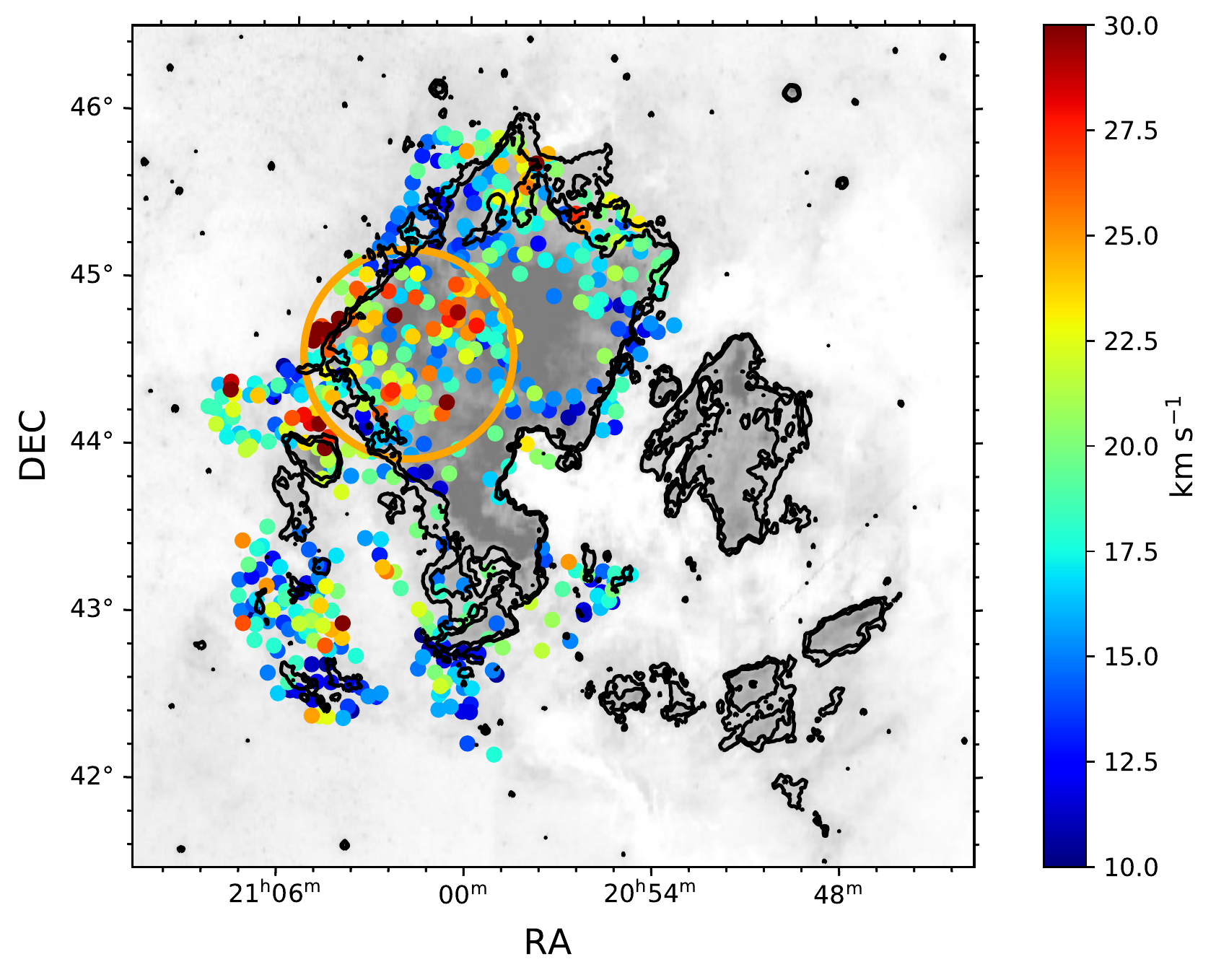}}
	\centerline{ (d)FWHMs of [\ion{N} {ii}] emission lines of NGC 7000}
\end{minipage}		
		\caption{ The spatial distributions of FWHMs of (a) H${\alpha}$ and (b) [\ion{N} {ii}] emission lines, and
		(c) (d) the close view for NGC 7000 region.
		The orange circle indicates the so called `wider FWHM region', and the four dashed ovals mark widest FWHM areas.
		The color bar shows the values of FWHMs.}
\label{Fig:FWHM}
\end{figure}

The FWHMs can provide us with information such as temperature, gas pressure, and star rotational velocity \citep{1895ApJ.....2..251M}. The FWHMs allow us to learn more about nebular properties.
Figure\ref{Fig:FWHM} shows the spatial distributions of FWHMs. Figure\ref{Fig:FWHM} (a) and (b) are the full field of view distributions for the W80 Region of H$\alpha$ and [\ion {N}{ii} ], respectively. Figure\ref{Fig:FWHM} (c) and (d) show a  close view for NGC 7000 region. The FWHMs in the regions covered by bright nebulae are narrow, mainly below 20 km s$^{-1}$. In the Middle Region, which is not covered by bright nebulae, we have discovered that the FWHMs are generally wider and the values over 30 km s$^{-1}$. In particular, Figure\ref{Fig:FWHM} (a) shows the blue dashed ovals in the Middle Region have the widest FWHMs, reaching 40 $\sim$ 50 km s$^{-1}$. We have labeled `M1' to `M4' nominating the blue dashed ovals from north to south in Figure \ref{Fig:FWHM} (a), in which the position and direction of the `M3' are consistent with the larger RV `jet feature'.

In addition, we have noticed a `wider FWHM region' in the eastern part of bright nebula NGC 7000 (see the orange circle in Figure\ref{Fig:FWHM}), shown by both H$\alpha$ and [\ion{N} {ii}] emission lines. The region has higher FWHM values than the other parts in NGC 7000. Figure\ref{Fig:FWHM} (c) and (d) display the close view for only NGC 7000 region, and show more clearly with only the 529 data points associated. This 'wider FWHM region' in fact coincides with the location of the Col428 cluster, and we discuss more about the association in Section \ref{subsection:4.3}.

\begin{figure}[!htp]
  \centering
  \includegraphics[width=0.7\textwidth,angle=0]{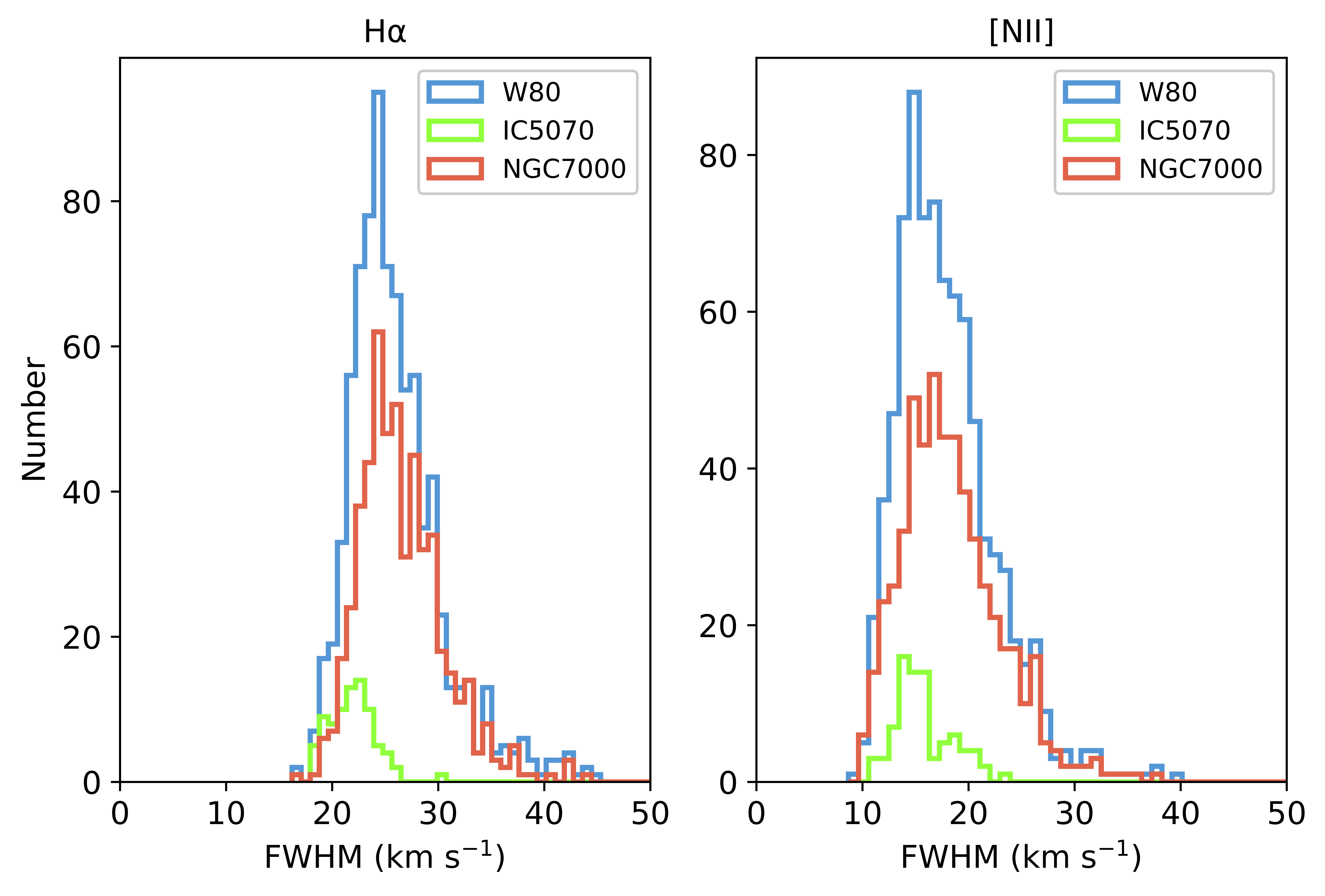}
		\caption{The histograms of FWHMs for H$\alpha$ and [\ion {N}{ii}] emission lines. }
  \label{Fig:fwhm_ha}
  \end{figure}

We present two sets of histograms to investigate the FWHM distributions. The histograms for the three regions are displayed in Figure\ref{Fig:fwhm_ha};
the left panel is NGC 7000, the middle panel is IC 5070, and the right panel is the W80 complex. 
The FWHMs of the IC 5070 region are the narrowest, with a peak for H${\alpha}$ emission lines at 21.9 km s$^{-1}$, and a peak for [\ion{N} {ii}] emission lines at 15.8 km s$^{-1}$.
In the NGC 7000 and W80 complex, the FWHM distributions of the H${\alpha}$ and [\ion{N} {ii}]emission lines are nearly identical, with peaks at 26 km s$^{-1}$ and 18 km s$^{-1}$, respectively.

\begin{figure}[!htp]
  \centering
  \includegraphics[width=0.7\textwidth,angle=0]{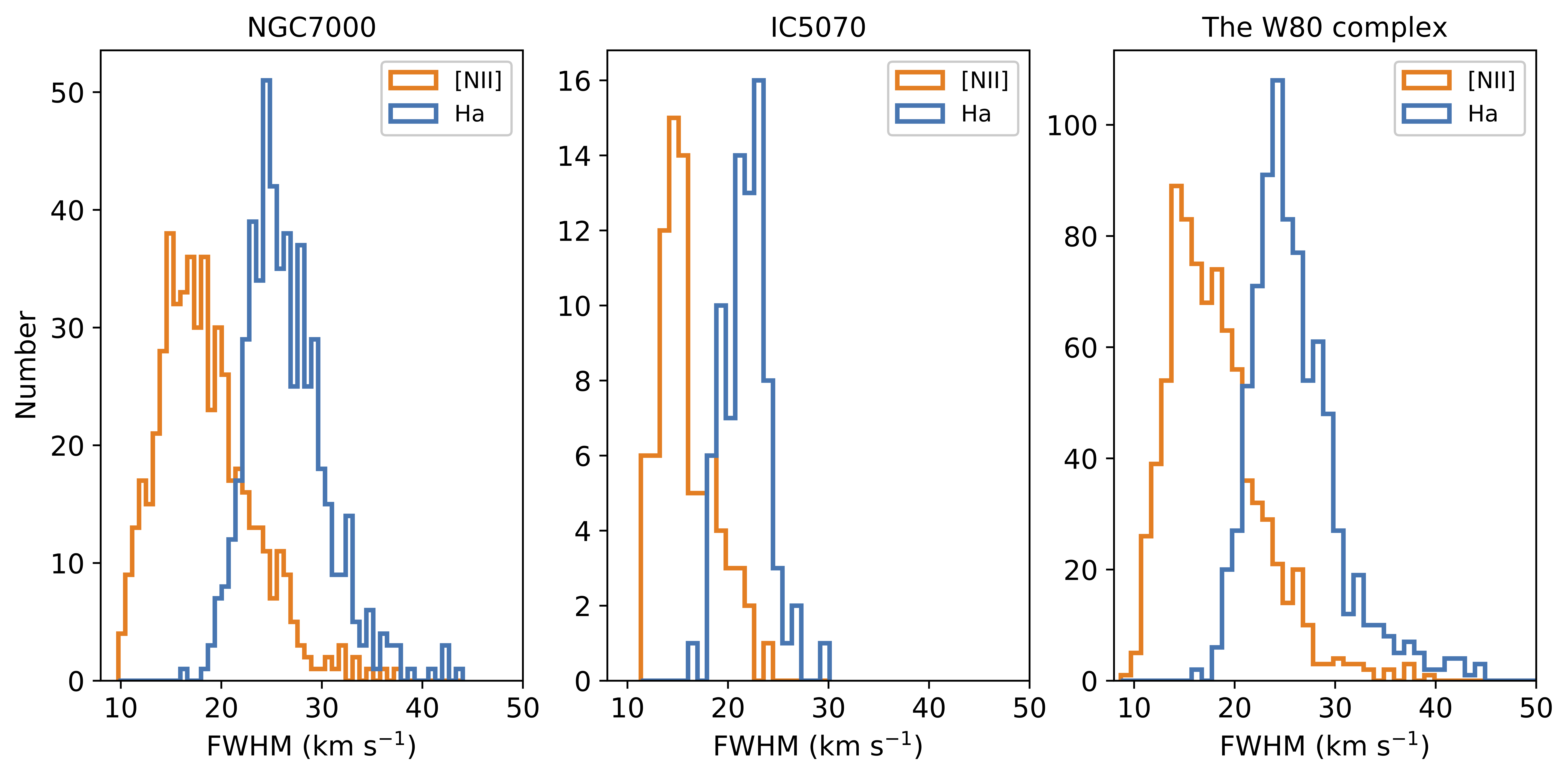}
	  \caption{The histograms of FWHMs for three regions. }
  \label{Fig:fwhm_NGC 7000}
  \end{figure}

Figure~\ref{Fig:fwhm_NGC 7000} shows two FWHM histograms for H${\alpha}$ and [\ion{N} {ii}] emission lines; the left panel is H${\alpha}$ and the right is the [\ion{N} {ii}] emission lines. The FWHMs of [\ion{N} {ii}] for NGC 7000, IC 5070, and W80 complex are 18.4 km s$^{-1}$, 15.8 km s$^{-1}$, and 18.1 km s$^{-1}$. The FWHMs of H${\alpha}$ lines are distributed in a range of 20$\sim$40 km s$^{-1}$, and FWHMs of [\ion{N} {ii}] lines in a range of 10$\sim$30 km s$^{-1}$; the FWHMs of H${\alpha}$ are about 10 km s$^{-1}$ larger than those of [\ion{N} {ii}]. We may understand the differences for heavier nitrogen than hydrogen. 

\subsection{The electron density $(n_{\mathrm{e}})$}
The electron density $n_{\mathrm{e}}$ can be calculated using a pair of nearly metastable lines. There is seldom a collision between ions and electrons in thin media, therefore the line ratio is exclusively determined by the Einstein coefficient. 
\cite{1970MNRAS.148..367S} have used line ratios as an indicator for density, to calculate electron densities in planetary nebulae. \cite{1970A&A.....9..175D} has used [\ion{S} {ii}] line or [\ion{O} {ii}] Line as a parameter to calculate the electron densities in \ion{H} {ii} regions.
We have employed the updated formula (see equation \ref{eq:R} and \ref{eq:ne}) proposed by \cite{2014A&A...561A..10P} to calculate the electron densities by using [\ion{S} {ii}] $\lambda\lambda$ 6716 \AA, 6731 \AA \ intensity ratios.

\begin{equation}
R=\frac{I_{6716}}{I_{6732}}
\label{eq:R}
\end{equation}

\begin{equation}
\begin{aligned}
\log \left(n_{\mathrm{e}}\right) &=0.0543 \tan (-3.0553 R+2.8506) \\
&+6.98-10.6905 R+9.9186 R^{2}-3.5442 R^{3}
\end{aligned}
\label{eq:ne}
\end{equation}

where $I_{6716}$ and $I_{6732}$ are the intensities of $\lambda$ 6716 \AA \  and $\lambda$ 6732 \AA \ lines, and R is the ratio. 
With the electron temperature of 10 000 K and the constraint of R \textless \ 1.42 (\cite{2014A&A...561A..10P}), we obtain an average $n_{\mathrm{e}}$ of 29.7 cm$^{-3}$ in the NGC 7000 with 121 spectral points, and 40.0 cm$^{-3}$ in the IC 5070 with 22 data points. 
For the whole W80 complex, the average $n_{\mathrm{e}}$ is estimated to be 43.2 cm$^{-3}$, with 171 spectral points. 
\cite{2019ApJ...880...16K} have obtained the electron densities of ionized nebulae in a range of 40-50000 cm$^{-3}$, with a new self-consistent [\ion{S}{ii}] doublet model. The electron densities of W80 complex could be around lower values to be probed by [\ion{S}{ii}] doublet. 

\section{Discussion}
\label{sect:discussion}
\subsection{Emission line measurements in the W80 Region}
\label{subsection:4.1}
The large FoV of LAMOST observations provides a powerful tool to reveal large-scale variations and unique structures in nebulae. This section summarizes the spectral observations to the W80 region with statistics for each regions and some structural features. The average values of RVs, relative intensities, and FWHMs for each corresponding regions are presented in Table~\ref{Tab:rotatedsample}. 

\begin{table}[!htp]
		\caption[]{ Emission line measurements in the W80 Region}\label{Tab:rotatedsample}
\begin{tabular}{llcccccccc}
\hline
\multicolumn{2}{l}{\multirow{2}{*}{Region}} & \multicolumn{3}{c}{\multirow{2}{*}{RVs(km s$^{-1}$)}} & \multicolumn{3}{c}{\multirow{2}{*}{revative intensity}} & \multicolumn{2}{c}{\multirow{2}{*}{FWHM (km s$^{-1}$)}} \\
\multicolumn{2}{l}{} & \multicolumn{3}{c}{} & \multicolumn{3}{c}{} & \multicolumn{2}{c}{} \\
\multicolumn{2}{l}{\multirow{2}{*}{}} & \multirow{2}{*}{H$\alpha$} & \multirow{2}{*}{[\ion{N}{ii}]} & \multirow{2}{*}{[\ion{S}{ii}]} & \multirow{2}{*}{H$\alpha$} & \multirow{2}{*}{[\ion{N}{ii}] } & \multirow{2}{*}{[\ion{S}{ii}] } & \multirow{2}{*}{H$\alpha$} & \multirow{2}{*}{[\ion{N}{ii}]} \\
\multicolumn{2}{l}{} &  &  &  &  &  &  &  &  \\ \hline
\multicolumn{2}{l}{\multirow{2}{*}{W80 complex}} & \multirow{2}{*}{-16.7$\pm$0.1} & \multirow{2}{*}{-16.2$\pm$0.2} & \multirow{2}{*}{-14.2$\pm$0.3} & \multirow{2}{*}{21.0} & \multirow{2}{*}{8.4} & \multirow{2}{*}{3.8} & \multirow{2}{*}{26.2$\pm$ 2.0} & \multirow{2}{*}{17.8$\pm$ 3.1} \\
\multicolumn{2}{l}{} &  &  &  &  &  &  &  &  \\
\multicolumn{2}{l}{\multirow{2}{*}{IC 5070}} & \multirow{2}{*}{-20.3$\pm$0.1} & \multirow{2}{*}{-20.1$\pm$0.1} & \multirow{2}{*}{-17.1$\pm$0.2} & \multirow{2}{*}{29.8} & \multirow{2}{*}{9.2} & \multirow{2}{*}{3.8} & \multirow{2}{*}{26.6$\pm$ 1.8} & \multirow{2}{*}{16.5$\pm$ 3.1} \\
\multicolumn{2}{l}{} &  &  &  &  &  &  &  &  \\
\multicolumn{2}{l}{\multirow{2}{*}{NGC 7000}} & \multirow{2}{*}{-16.1$\pm$0.1} & \multirow{2}{*}{-15.7$\pm$0.2} & \multirow{2}{*}{-14.2$\pm$0.3} & \multirow{2}{*}{24.4} & \multirow{2}{*}{7.5} & \multirow{2}{*}{4.6} & \multirow{2}{*}{26.7$\pm$ 1.9} & \multirow{2}{*}{18.0$\pm$ 2.9} \\
\multicolumn{2}{l}{} &  &  &  &  &  &  &  &  \\
\multicolumn{2}{l}{\multirow{2}{*}{-curved   feature}} & \multirow{2}{*}{-18.9$\pm$ 0.2} & \multirow{2}{*}{-18.7$\pm$0.2} & \multirow{2}{*}{-17.1$\pm$0.3} & \multirow{2}{*}{30.0} & \multirow{2}{*}{13.0} & \multirow{2}{*}{5.8} & \multirow{2}{*}{29.2$\pm$ 2.0} & \multirow{2}{*}{21.1$\pm$2.9} \\
\multicolumn{2}{l}{} &  &  &  &  &  &  &  &  \\
\multicolumn{2}{l}{\multirow{2}{*}{-wider FWHM}} & \multirow{2}{*}{-18.6$\pm$ 0.2} & \multirow{2}{*}{-18.3$\pm$0.2} & \multirow{2}{*}{-16.8$\pm$0.2} & \multirow{2}{*}{37.7} & \multirow{2}{*}{17.1} & \multirow{2}{*}{7.6} & \multirow{2}{*}{28.9$\pm$ 2.2} & \multirow{2}{*}{20.9$\pm$3.2} \\
\multicolumn{2}{l}{} &  &  &  &  &  &  &  &  \\
\multicolumn{2}{l}{\multirow{2}{*}{LBN 391}} & \multirow{2}{*}{-13.6$\pm$0.2} & \multirow{2}{*}{-12.7$\pm$0.3} & \multirow{2}{*}{-11.1$\pm$0.4} & \multirow{2}{*}{8.2} & \multirow{2}{*}{3.6} & \multirow{2}{*}{1.9} & \multirow{2}{*}{26.7$\pm$ 1.8} & \multirow{2}{*}{16.5$\pm$ 3.1} \\
\multicolumn{2}{l}{} &  &  &  &  &  &  &  &  \\
		\multicolumn{2}{l}{\multirow{2}{*}{-Northern (LBN-N)}} & \multirow{2}{*}{-9.2$\pm$ 0.1} & \multirow{2}{*}{-8.2$\pm$0.2} & \multirow{2}{*}{-6.8$\pm$0.4} & \multirow{2}{*}{12.8} & \multirow{2}{*}{5.3} & \multirow{2}{*}{2.5} & \multirow{2}{*}{26.1$\pm$ 1.8} & \multirow{2}{*}{16.5$\pm$3.1} \\
\multicolumn{2}{l}{} &  &  &  &  &  &  &  &  \\
		\multicolumn{2}{l}{\multirow{2}{*}{-Southern (LBN-S)}} & \multirow{2}{*}{-16.6$\pm$ 0.2} & \multirow{2}{*}{-15.8$\pm$0.2} & \multirow{2}{*}{-14.2$\pm$0.5} & \multirow{2}{*}{8.1} & \multirow{2}{*}{3.5} & \multirow{2}{*}{1.7} & \multirow{2}{*}{24.9$\pm$ 2.1} & \multirow{2}{*}{15.5$\pm$3.6} \\
\multicolumn{2}{l}{} &  &  &  &  &  &  &  &  \\
\multicolumn{2}{l}{\multirow{2}{*}{-jet feature}} & \multirow{2}{*}{-18.8$\pm$ 0.3} & \multirow{2}{*}{-16.4$\pm$0.4} & \multirow{2}{*}{-15.6$\pm$0.7} & \multirow{2}{*}{5.3} & \multirow{2}{*}{2.0} & \multirow{2}{*}{1.4} & \multirow{2}{*}{37.0$\pm$ 1.7} & \multirow{2}{*}{30.0$\pm$2.8} \\
\multicolumn{2}{l}{} &  &  &  &  &  &  &  &  \\
\multicolumn{2}{l}{\multirow{2}{*}{Middle   Region}} & \multirow{2}{*}{-11.2$\pm$0.4} & \multirow{2}{*}{-10.2$\pm$0.5} & \multirow{2}{*}{-8.8$\pm$0.8} & \multirow{2}{*}{3.3} & \multirow{2}{*}{1.6} & \multirow{2}{*}{1.1} & \multirow{2}{*}{35.8$\pm$ 1.9} & \multirow{2}{*}{25.3$\pm$ 3.3} \\
\multicolumn{2}{l}{} &  &  &  &  &  &  &  &  \\
\multicolumn{2}{l}{\multirow{2}{*}{-Northern (M1)}} & \multirow{2}{*}{-7.2$\pm$ 0.3} & \multirow{2}{*}{-6.4$\pm$0.4} & \multirow{2}{*}{-6.0$\pm$0.5} & \multirow{2}{*}{3.9} & \multirow{2}{*}{1.9} & \multirow{2}{*}{1.5} & \multirow{2}{*}{36.6$\pm$ 1.9} & \multirow{2}{*}{24.2$\pm$3.0} \\
\multicolumn{2}{l}{} &  &  &  &  &  &  &  &  \\
		\multicolumn{2}{l}{\multirow{2}{*}{-Southern (M4)}} & \multirow{2}{*}{-13.0$\pm$ 0.4} & \multirow{2}{*}{-12.1$\pm$0.7} & \multirow{2}{*}{-10.4$\pm$1.0} & \multirow{2}{*}{3.0} & \multirow{2}{*}{1.4} & \multirow{2}{*}{0.9} & \multirow{2}{*}{36.8$\pm$ 2.2} & \multirow{2}{*}{26.9$\pm$4.1} \\
\multicolumn{2}{l}{} &  &  &  &  &  &  &  &  \\ \hline
\end{tabular}
\end{table}

For the W80 Region on the whole, the bright nebulae in NGC 7000, IC 5070 and LBN935 systematically possess larger RV values than the Middle Region; the mean RV of H$\alpha$ emission lines in the W80 complex is -16.7 km s$^{-1}$, and -11.2 km s$^{-1}$ in the Middle Region. Considering the large scale of detection, we speculate that the systematical difference of the RVs could indicate an overall property of the local molecular clouds; in fact, the filamentary molecular clouds in this region are just distributed along the N-S direction \citep{2014AJ....147...46Z, 2021AJ....161..229K}.

The overall spatial distribution of FWHMs also shows the bright nebular regions have narrower FWHMs than the Middle Region; the mean FWHM of H$\alpha$ lines in the W80 complex is 26.2 km s$^{-1}$, and 35.8 km s$^{-1}$ in the Middle Region. For the bright nebulae the emission lines are observing a relatively short physical scale of the \ion{H}{ii} regions, while the weak line emissions in the Middle Region may indicate optical thin and so that longer distances along the line of sight. More velocity components along the long path could easily lead to relatively larger FWHMs.

As for different emission lines, the RVs of the H$\alpha$ emission lines are basically the same as those of the [\ion{N}{ii}] emission lines and slightly larger than those of the [\ion{S}{ii}] emission lines.
The mean RVs of the H$\alpha$ and [\ion{N}{ii}] emission lines for the W80 complex are both about -16 km s$^{-1}$, and -14.2 km s$^{-1}$ of the [\ion{S}{ii}] emission lines.
The FWHMs of H$\alpha$ emission lines are generally wider than the [\ion{N}{ii}] emission lines. The FWHM of the two emission lines for the W80 complex are 26.2 and 17.8 km s$^{-1}$, respectively.

In the NGC 7000, the `curved feature' shows a RV value about 2 km s$^{-1}$ larger than the entire NGC 7000 region, and the `wider FWHM region' has the FWHM about 3 km s$^{-1}$ wider than other regions.

The southern and northern regions of the LBN 391 possess obviously different radial velocities,  with the RVs of H$\alpha$ emission lines of -16.6 km s$^{-1}$ and -9.2 km s$^{-1}$ , respectively. The RV difference of the two regions is about 7 km s$^{-1}$, but their FWHMs are quite similar.

Several components in the Middle Region are also worth noting; the mean RV in the northern region (M1) is -7.2 km s$^{-1}$ and the mean FWHM is 36.6 km s$^{-1}$, and -13.0 km s$^{-1}$ and 36.8 km s$^{-1}$ in the southern region (M4). The RV difference are about 6 km s$^{-1}$ between the southern and northern components, and their FWHMs are similar.

\begin{table}[!htp]
 \caption[]{The RVs and FWHMs for the H$\alpha$ emission line in the literature }\label{Tab:bijiaoRV}
\begin{threeparttable}
 \begin{tabular}{ccccccc}
 \hline
literature & \multicolumn{2}{c}{IC 5070} & \multicolumn{2}{c}{NGC 7000} & \multicolumn{2}{c}{W80} \\
 & RVs (km s$^{-1}$) & FWHM (km s$^{-1}$)  & RVs (km s$^{-1}$)  & FWHM (km s$^{-1}$)  & RVs (km s$^{-1}$)  & FWHM (km s$^{-1}$)   \\ 		 \hline
PM73 \tnote{1}  & -16.4 & 28.7  & -15.7 & $\sim$ & -16.8 & 23.3 \\
F80 \tnote{2} & -14.8 $\pm$ 4.1 & 37.8$\pm$ 12.7& -15.4 $\pm$ 3.7 & 22.5 $\pm$ 3.6 & -15.1 $\pm$ 5.5 & 28.6$\pm$ 0.6  \\
DEFPOS \tnote{3} & -8.8 $\pm$ 0.9 & 49.7 $\pm$ 2.8 & -7.4 $\pm$ 0.6 & 50.2 $\pm$ 1.8 & -7.8 $\pm$ 0.7 & 50.0 $\pm$ 2.1 \\
MRS-N  & -20.1 $\pm$  0.1 & 26.6 $\pm$ 1.8 & -15.8 $\pm$  0.1 & 26.7 $\pm$ 1.9 & -16.4 $\pm$ 0.1 & 26.2 $\pm$  2.0 \\ \hline
 \end{tabular}
\begin{tablenotes}
\item[1]\cite{1973MNRAS.165..381P};
\item[2]\cite{1983apj...269..164f};
\item[3]\cite{2013RAA....13..547A}.
 \end{tablenotes}
\end{threeparttable}
\end{table}

Table \ref{Tab:bijiaoRV} compares the results of the previous literature mentioned in the introduction with this work. The big differences appear only in the DEFPOS observations \citep{2013RAA....13..547A}.  \cite{2011NewA...16..485A} have statistically analyzed the spectra of 10 nebulae with DEFPOS, and stated that the FWHMs in DEFPOS are 10 km s$^{-1}$ wider than those in other literature, and the radial velocity of DEFPOS has a deviation of 3 km s$^{-1}$.

\subsection{Features revealed by radial velocity }
\label{subsection:4.2}
Based on the results of RV spatial distributions (see Section 3.2), we have discovered a `curved feature' in the eastern extension of the NGC 7000, and a `jet feature' extended from the LBN 391 nebula. Both the larger RV features are located beyond their bright nebulae and appear as narrow strips with  uniform velocity components. 

We have noticed that the larger RVs appear in the regions associated with outflow activities in bright nebula IC 5070 and dark cloud L935.
\cite{2003AJ....126..893B} and \cite{2011A&A...528A.125A} have discovered many HH objects to the IC 5070, and \cite{2014AJ....148..120B} have also detected the MHOs to the L935; 
the comparison of these multi-band observations may provide a hint for understanding the larger RV structures. 
All the features require more observations to figure out in the infrared and submillimeter wavelengths.

There are other notable features in the LBN 391 nebula. The position and direction of the `jet feature' in the radial velocity map are consistent with the widest FWHM region (`M3') in the FWHM map, and this `jet feature' appears at the boundary of the velocities differences between the northern and southern regions of LBN 391. But the `curved feature' extended to the east of the NGC 7000 does not show wider FWHMs. All of these newly discovered characteristic structures require more observations and verification, and more comprehensive spectral observations of the LBN 391 nebula have been planned.

\subsection{The `wider FWHM region' of NGC 7000}\label{subsection:4.3}
Based on the observational results of the FWHM distributions in Section 3.3, we have noticed an anomalous `wider FWHM region' in the eastern part of NGC 7000, consistent with the location of Col428 cluster.
The Col428 stellar cluster has been first discovered by \cite{1931AnLun...2....1C} and classified as an open cluster. 
But later, \cite{2007BaltA..16..349L} have figured out that the Col428 is probably not a stellar cluster demonstrated through the distance and color-magnitude diagram, 
and \cite{2011ApJS..193...25R} further pointed out that Col428 could not be a real young cluster because of a low disk fraction; both of them have proposed that the Col428 could not be taken as an independent entity but be a `window' in the molecular cloud. 
In this scenario, the `wider FWHM region' may be understood that we are observing an optical thin region, and more components deeply detected in the \ion{H}{ii} region contribute to the line emissions. 

For understanding the cause of `wider FWHM region', we have tried to  calculate further the gas temperature and non-thermal motion velocity. Based on the measured FWHMs towards \ion{H}{ii} regions, \cite{1977ApJ...211..115R} have established a relationship between gas temperature and non-thermal motion (see equations \ref{eq:gongshiha} and \ref{eq:gongshiSII}). 

\begin{equation}
	T(\mathrm{~K})=23.5 W_{\mathrm{H}}^{2}\left(1-\frac{W_{\mathrm{N}}^{2}}{W_{\mathrm{H}}^{2}}\right)
\label{eq:gongshiha}
\end{equation}

\begin{equation}
		V\left(\mathrm{~km} \mathrm{~s}^{-1}\right)=1.04 W_{\mathrm{N}}\left(1-0.071 \frac{W_{\mathrm{H}}^{2}}{W_{\mathrm{N}}^{2}}\right)^{1 / 2}
\label{eq:gongshiSII}
\end{equation}
where $W_{\mathrm{H}}$ and $W_{\mathrm{N}}$
are the FWHMs of H${\alpha}$ and [\ion{N} {ii}] emission lines, $T$ is gas temperature, and $V$ is non-thermal velocity. 

The non-thermal motion velocities and gas temperatures have been calculated for the spectral points inside and outside the `wider FWHM region', respectively. We note that, for the 195 statistical spectral points, 76 inside and 119 outside the `wider FWHM region', the FWHM uncertainties are 1.6 km s$^{-1}$ for H${\alpha}$, and 2.5 km s$^{-1}$ for [\ion{N} {ii}] emission lines, and all the calculated $T$ and $V$ are selected to be greater than 3 times the standard deviations. The uncertainties of $T$ and $V$ are calculated by following the usual error propagation equation. 
The results show that the `wider FWHM region' has a mean non-thermal velocity of 20.5 $\pm$ 2.9 km s$^{-1}$ and a mean temperature of 10269 $\pm$ 3551 K, while the outside region of 16.8 $\pm$ 2.7 km s$^{-1}$ and 9134 $\pm$ 2794 K. 
For the differences of non-thermal motion velocities and gas temperature are both not much bigger than the uncertainties, it seems hard to be sure which could  responsible for the `wider FWHM region'. 

\subsection{Features of the W80 complex }\label{subsection:4.4}
\begin{figure}[!htp]
  \centering
	    \includegraphics[width=0.7\textwidth, angle=0]{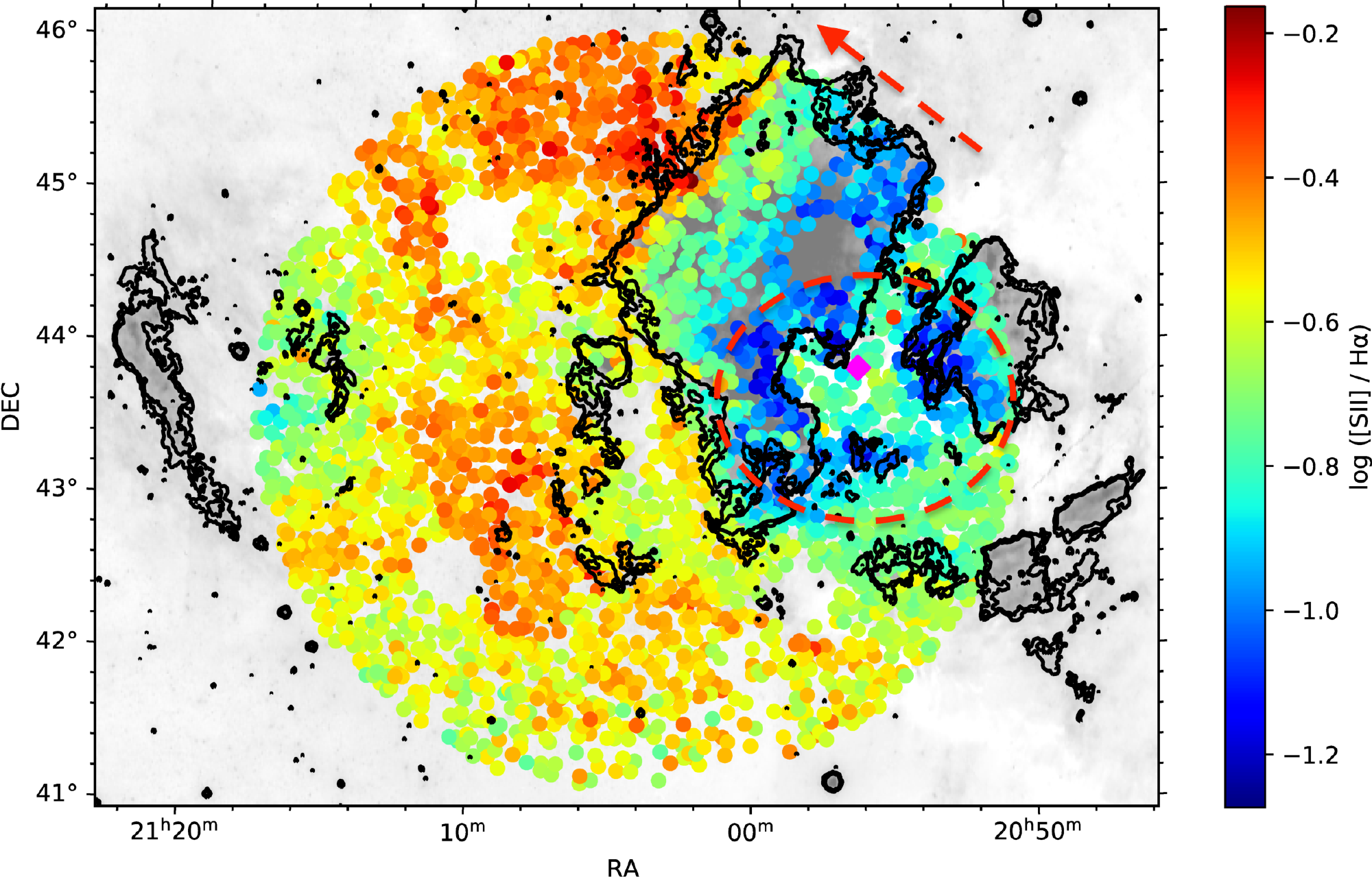}
	    \caption{The distributions of the log ([\ion{S}{ii}] / H${\alpha}$) ratios. The color bar shows the log ([\ion{S}{ii}] / H${\alpha}$) ratios in the W80 region. The red dashed circle marks the ring shape with a low ratios around L935, and the red arrow points to the direction of  increasing gradient in the NGC 7000 nebula. }
\label{Fig:tidu}
\end{figure}

The H$\alpha$ emission represents photoionization with dense ultraviolet photons in \ion{H}{ii} regions, and the [\ion{S}{ii}] forbidden line emission is commonly observed at the outer edge of nebulae. The nebular structures can be further revealed by comparing the intensities of [\ion{S}{ii}] and the H$\alpha$  emission lines.

Figure~\ref{Fig:tidu} presents the distributions of the log ([\ion{S}{ii}] / H${\alpha}$) ratios in the W80 region. There are two interesting features associated with the W80 complex. 
Firstly, the minimum ratios are found to distribute around the `W80 bubble', showing as a ring shape. The `W80 bubble' appears as an ionized bubble observed at radio wavelengths \citep{1958BAN....14..215W}, and its ionizing source is an O3.5 star (2MASS J20555125+4352246) \citep{2005a&a...430..541c}.

A low [\ion{S}{ii}]/H${\alpha}$ intensity ratio indicates relatively strong H${\alpha}$ emissions, whereas a high ratio means relatively strong [\ion{S}{ii}] emissions. 
The ultraviolet photons from the central excitation source of a massive star ionize the surrounding gas to form stronger H${\alpha}$ emission lines in the `W80 bubble' region, which results in a significant decrease of [\ion{S}{ii}]/H${\alpha}$ ratios. 
Moreover, due to the occlusion of the high-density molecular cloud in the foreground, it shows a ring-shaped distribution centered on the excitable O-type star. Therefore, the ring-shape with low [\ion{S}{ii}]/H$\alpha$ ratios revealed by our spectral observations in the optical actually supports the `W80 bubble' structure observed by radio wavelengths.

In the western part of the NGC 7000, there is also a region with low [\ion{S}{ii}]/H$\alpha$ ratios, associated with the N6997, a dense young stellar cluster. The feature could provide evidence of abundant ultraviolet photons to the direction of N6997. 

Secondly, an increasing gradient of [\ion{S}{ii}]/H$\alpha$ ratios can be found from southwest to northeast in the NGC 7000 region. 
The increase of [\ion{S}{ii}]/H$\alpha$ ratios, indicates enhancement of [\ion{S}{ii}] emission and the weakening of photoionization. We note that the strongest [\ion{S}{ii}] emissions at the northeastern edge of the NGC 7000 (Figure \ref{Fig:RE}) are consistent with the highest [\ion{S}{ii}]/H$\alpha$ ratios in Figure ~\ref{Fig:tidu}. The variations of the [\ion{S}{ii}]/H$\alpha$ ratios have been commonly observed, both in the ionization structure \citep{2019ApJ...880...16K} and in diffuse ionized gas \citep{2002ApJ...572..823O}. It is generally acknowledged that the  hard ionizing radiation field escaped out of \ion{H}{ii} regions could produce strong [\ion{S}{ii}] line emissions \citep{2019ApJ...880...16K}.

In addition, high [\ion{S}{ii}]/H$\alpha$ ratios are prominent in the Middle Region. \cite{2009ApJ...704..842B} have pointed out that [\ion{S}{ii}]/H$\alpha$ is observed to be greatly enhanced in diffuse ionized gas, and provided a mean [\ion{S}{ii}]/H$\alpha$ ratio of 0.34, that is, -0.46 in log  [\ion{S}{ii}]/H$\alpha$. The ratio seems just consistent with the measurements in the Middle Region in Figure \ref{Fig:tidu}.

\section{Summary}
\label{sect:summary}
The large-scale spectral observations to the \ion{H}{ii} regions are important for us to understand the star formation process and help study the environment of the whole W80 Region. With the LAMOST higher spectral resolution and the wider FoV of 20 square degrees, we have built an almost complete spectral sample of the W80 emission nebulae. The results of this study are summarized as follows.

(1) In the W80 Region, based on the 3800 spectra by the LAMOST MRS-N Survey, we obtain a total of 2982  nebular spectra with high enough signal-to-noise ratios, establishing the largest sample of spectral data up to date. 
We have measured the relative intensities, the RVs, and the FWHMs of the H${\alpha}$, [\ion{N} {ii}], and [\ion{S} {ii}] emission lines for the W80 Region, and have estimated as far as possible the electron densities for the W80 complex.

(2) The spatial distributions of relative intensities in the W80 Region show that 
the strongest line emissions are found in the bright nebular regions, like NGC 7000, IC 5070, and LBN 391; the weak line emissions also truly exist in the Middle Region, where no bright nebulae detected in the wide-band optical observations. The strongest H$\alpha$ emission in the NGC 7000 concentrates mainly in the central part, while the stronger [\ion{N}{ii}] and [\ion{S}{ii}] emissions tend to be distributed at the northeastern edge.

(3) The spatial distributions of radial velocities in the W80 Region show overall radial velocity differences. The bright nebulae display radial velocities moving towards us, with the velocities mainly around -25 to -15 km s$^{-1}$. The Middle Region has radial velocities ranging from -10 to 0 km s$^{-1}$, and some velocity components in the northern part are far away from us, with velocities ranging from 0 to +5 km s$^{-1}$.

(4) The spatial distributions of FWHMs in the W80 Region also show overall differences, 
revealing that the FWHMs in the areas covered by the bright nebulae are narrower, mainly below 20 km s$^{-1}$; the FWHMs in the Middle Region are wider, above 30 km s$^{-1}$. In the Middle Region there exist several parts with the widest FWHM values ranging from 40 to 50 km s$^{-1}$. 

(5) The large-scale spectral observations to the W80 Region also reveal some unique structural features. A `curved feature' to the east of the NGC 7000, and a `jet feature' to the west of the LBN 391 are detected to be showing with larger approaching radial velocities. A `wider FWHM region' is identified in the eastern part of the NGC 7000. The variations of [\ion{S}{ii}] / H${\alpha}$ ratios display a gradient from southwest to northeast in the NGC 7000 region, and manifest a ring shape around the `W80 bubble' in the L935. 

The new scientific findings by the MRS-N provide useful information for researching the large-scale properties of the \ion{H}{ii} regions. For complete data sets for the W80 region, we will conduct further spectral observations, and combine multi-band observations to investigate in detail the structural features.

\begin{acknowledgements}
The authors thank sincerely the anonymous referee for the great help to improve this paper. We also want to thank Xingchen Liu for the warmhearted help with data reduction and Hongchi Wang, Zhibo Jiang, Shaobo Zhang and Zhiwei Chen for scientific discussions. 

This project is supported by the National Natural Science Foundation of
 China (Grant Nos.  12073051, 11973004, 12090040, 12090041, 11733006, 11403061, 11903048, U1631131, 11973060, 12090044, 12073039, 11633009, U1531118), and the Key Laboratory of Optical Astronomy, National Astronomical Observatories, Chinese Academy of Sciences, and the Key Research Program of Frontier Sciences, CAS (Grant No. QYZDY-SSW- SLH007).

C.-H. Hsia acknowledges the supports from the Science and Technology Development Fund, Macau SAR (file No. 0007/2019/A) and Faculty Research Grants of the Macau University of Science and Technology (No. FRG- 19-004-SSI).

Guoshoujing Telescope (the Large Sky Area Multi-Object Fiber Spectroscopic Telescope LAMOST) is a National Major Scientific Project built by the Chinese Academy of Sciences. Funding for the project has been provided by the National Development and Reform Commission. LAMOST is operated and managed by the National Astronomical Observatories, Chinese Academy of Sciences.
\end{acknowledgements}

\label{lastpage}
\bibliographystyle{raa}
\bibliography{sample}

\end{document}